\documentclass[12pt, a4paper]{article}

 \usepackage[bottom]{footmisc}

\usepackage{graphicx}
\usepackage{amsthm}   
\usepackage{amsmath} 
\usepackage{amssymb}  
\usepackage{mathrsfs} 
\usepackage{stmaryrd} 
\usepackage{txfonts} 

\usepackage{hyperref}  

\usepackage[capitalize]{cleveref}

\usepackage{color}
\RequirePackage[dvipsnames,usenames]{xcolor}

\newcommand{\ba}{\begin{eqnarray}}
\newcommand{\ea}{\end{eqnarray}}

\newcommand{\eq}[1]{\begin{align}#1\end{align}}

\newcommand{\K}{{\overline{K}}}
\newcommand{\II}{{\mathcal{I}}}

\newcommand{\AAA}{{\mathcal{A}}}
\newcommand{\BBB}{{\mathcal{B}}}
\newcommand{\MMM}{{\mathcal{M}}}

\newcommand{\R}{{\mathbb{R}}}

\newcommand{\NN}{\mathcal{N}}

\newcommand{\oPi}{{\overline{\Pi}}}
\newcommand{\opi}{{\overline{\pi}}}
\newcommand{\oZ}{{\overline{Z}}}

\newcommand{\oW}{{\overline{W}}}
\newcommand{\oQ}{{\overline{Q}}}
\newcommand{\vx}{{\pmb{x}}}
\newcommand{\vX}{{\pmb{X}}}
\newcommand{\oo}{{\omega}}
\newcommand{\vP}{{\bold{P}}}
\newcommand{\ovP}{{\overline{\bold{P}}}}

\newtheorem{example}{Example}

\begin{document}


\title{Combining lower bounds on entropy production in complex systems with multiple interacting components}

\author{David H. Wolpert \\
 Santa Fe Institute, 1399 Hyde Park Road, Santa Fe, NM, 87501\\
%
%
Arizona State University, AZ, USA \\
Complexity Science Hub, Vienna, Austria \\
International Center for Theoretical Physics, Trieste, Italy \\
Albert Einstein Institute for Advanced Study, NY, USA \\
http://davidwolpert.weebly.com
}

\maketitle

\begin{abstract}
The past two decades have seen a revolution in statistical physics, generalizing it
to apply to 
systems of arbitrary size, evolving while arbitrarily far from equilibrium. Many of these new results are based on analyzing the dynamics of 
the entropy of a system that is evolving according to a Markov process. These results comprise a sub-field called ``stochastic thermodynamics''.
Some of the most powerful results in stochastic thermodynamics were traditionally 
concerned with single, monolithic systems, evolving by themselves, ignoring any internal structure of those systems.
In this chapter I review how in complex systems, composed of many interacting constituent systems, it is possible
to substantially strengthen many of these traditional results of stochastic thermodynamics. This is done by ``mixing and matching''
those traditional results, to each apply to only a subset of the interacting systems, thereby producing a more powerful
result at the level of the aggregate, complex system. 
\end{abstract}

\section{Introduction}

Starting around the turn of the millennium, non-equilibrium statistical physics began to undergo
a revolution. We can now analyze the thermodynamic behavior of many kinds of system, of arbitrary size, evolving while arbitrarily
far from thermodynamic equilibrium~\cite{seifert2012stochastic,van2015ensemble,peliti2021stochastic}.
This has given us a far more profound understanding of topics ranging from the thermodynamics of computation~\cite{wolpert_thermo_comp_review_2019}
to the thermodynamics of electronic circuits~\cite{freitas2020stochastic} to the thermodynamics of chemical reaction networks~\cite{esposito2020open}.

The starting point for many of these results is surprisingly simple.
Statistical physics concerns experimental scenarios where we have restricted information concerning
 the state of a system $x \in X$, which is quantified as a probability distribution 
over those states, $p_x(t)$. The sub-field of modern non-equilibrium statistical physics
known as ``stochastic thermodynamics'' is, at heart, simply the analysis of how the Shannon entropy of such a distribution
evolves as that distribution undergoes a continuous-time Markov chain (CTMC). 
For a countable state space, this means analyzing how the Shannon entropy of $p_x(t)$ evolves if $p_x(t)$ itself evolves according
to a linear differential equation,
\eq{
\dfrac{d p_x(t)}{dt} = \sum_{x'} K^{x'}_x(t) p_{x'}(t)
\label{eq:0}
}
where $K(t)$ is a stochastic rate matrix. 
(Note that the rate matrix can depend on time $t$.)

%
%

Analyzing the entropy of systems which are evolving according to~\cref{eq:0} has led to formulations of the 
second law of thermodynamics which apply even if the system 
is evolving while arbitrarily far out of thermal equilibrium~\cite{van2015ensemble,seifert2012stochastic}.
In particular, assume that the system is evolving according to \cref{eq:0} 
while coupled to a single (infinite) heat
bath at temperature $T$. Assume as well that the rate matrix is related to
the underlying Hamiltonian of the system via \textbf{local detailed balance} (LDB). (This is equivalent to assuming micro-reversibility
of the dynamics,
or viewed differently, it is equivalent to assuming thermodynamic consistency~\cite{seifert2012stochastic}.) Then if we 
apply one of the mentioned new analyses of the dynamics of the entropy, we get
\eq{
\dfrac{Q}{T} \le \Delta S
\label{eq:1}
}
where $Q$ is the total heat flow into the system from its heat bath during the dynamics,
and $\Delta S$ is the change in Shannon entropy of the system during the process. 

If LDB does not hold, \cref{eq:1} will not hold either, if we wish to interpret $Q$ as 
thermodynamic heat flow. However, for any rate matrix,
regardless of whether it obeys LDB, 
\eq{
\int_{t_i}^{t_f} dt  \sum_{x'} K^{x'}_x(t) p_{x'}(t) \ln \dfrac{K^{x}_{x'}(t)}{K^{x'}_{x}(t)} \,  \le \Delta S
\label{eq:2}
}
(for a process lasting from time $t_i$ to $t_f$). The quantity on the LHS of \cref{eq:2} is
called the total expected \textbf{entropy flow} (EF) into the system during the process. The difference
between the entropy change of the system (the RHS of \cref{eq:2}) and the EF is called the \textbf{entropy production} (EP),
written as $\sigma$. So \cref{eq:2} can be re-expressed as
\eq{
\sigma \ge 0
\label{eq:2a}
}
Crucially, the inequality \cref{eq:2a} holds for \textit{any} CTMC, even 
a CTMC that has no thermodynamic interpretation,
i.e., a CTMC which models a process that does not involve energy transduction. 
So \cref{eq:2a} applies to dynamic models of everything from stock markets to the evolution of the joint state
of an opinion network, so long as those models are CTMCs.

Stochastic thermodynamics goes far beyond \cref{eq:2a} however.
For example, phrased more carefully, \cref{eq:2a} concerns the expected value of EP during the CTMC, 
evaluated across all trajectories through the state space whose
relative probabilities are determined by the CTMC. The recently derived ``fluctuation theorems'' (FTs)~\cite{seifert2012stochastic,van2015ensemble,Bisker_2017,esposito2010three,wolpert_thermo_comp_review_2019}
take this further, by providing information about the full probability distribution of the value of the total EP generated over any 
time interval by a system that evolves according to a CTMC. 
As an illustration of their power, the FTs provide bounds strictly more powerful than 
the second law, leading them to be identified as the ``underlying
thermodynamics ... of time's arrow''~\cite{seif2020machine}. 

As another example of the powerful results of stochastic thermodynamics, note that in many experimental scenarios, 
while we are restricted in the information we have concerning 
the system's state, only having a distribution $p_x(t)$, we have some other information, in the form of conditions satisfied by the dynamics 
of the system. More formally, in the real world, not all rate matrices are possible, and we often have
\textit{a priori} information concerning what form the rate matrix of a given experimental system could have.
Recently \cref{eq:2a} has been strengthened, by adding non-negative terms to its RHS
that incorporate this kind of information concerning the dynamics. 

An important set of results of this type are known as the ``Thermodynamic Uncertainty relations''
(TURs)~\cite{horowitz_gingrich_nature_TURs_2019,liu2020thermodynamic,hasegawa2019fluctuation,koyuk2020thermodynamic,falasco2020unifying,barato2015thermodynamic}. Each TUR applies to a different set of processes, i.e., each one holds 
for a different set of constraints on the dynamics of the system. 
What unites the TURs is that each one provides a lower bound on the EP generated during a process
in terms of the statistical precision of \textit{any} stochastic current quantifying the net number of transitions among the
system states during that process.
So the closer the distribution of values of such a current is to a delta function --- loosely
speaking, the closer the current is to a deterministic process --- the greater the EP that
must be expended. Perhaps the most famous example of a TUR is the bound 
\eq{
\langle \sigma \rangle \ge \dfrac{2\langle J\rangle^2 }{{\mbox{Var}}(J)} 
\label{eq:ness_tur}
}
which applies to any system that is in a nonequilibrium stationary states (NESS).
I will refer to \cref{eq:ness_tur} as the  ``canonical TUR''~\cite{horowitz_gingrich_nature_TURs_2019}.

These two recent sets of results --- FTs and TURs --- are closely related. In particular, 
an FT can be used to derive a TUR~\cite{hasegawa2019fluctuation}.
%
In fact, recently there has been a veritable explosion of results related to FTs and TURs, which use aspects
of the EP during a process to bound other quantities of interest.
Examples of these new results include ``speed limit theorems'' (SLTs~\cite{shiraishi_speed_2018,zhang2018comment,okuyama2018quantum,van2020unified,gupta2020thermodynamic}),
``thermodynamic first passage bounds''~\cite{neri2017statistics,gingrich2017fundamental,falasco2020dissipation,roldan2015decision,neri2021dissipation}, 
``thermodynamic stopping time bounds''~\cite{falasco2020dissipation}, etc. 

Importantly, almost all of these results have considered physical system monolithically, without
taking into account how they decompose into smaller systems. However, 
we are often interested the thermodynamics of large, complex systems, that can be modeled as multiple interacting smaller systems.
Examples include a digital circuit, which decomposes into a set of interacting 
gates~\cite{wolpert_thermo_comp_review_2019,wolpert2018thermo_circuits}, and 
a cell, which decomposes into a set of many organelles and biomolecule species.

Recent research in stochastic thermodynamics
has started to investigate such aggregate, complex systems~\cite{sagawa2008second,sagawa2009minimal,parrondo2015thermodynamics,horowitz2014thermodynamics,barato_efficiency_2014,ito2013information,hartich_sensory_2016}.
Most of this research has considered the special case of bipartite 
processes~\cite{sagawa2008second,sagawa2009minimal,parrondo2015thermodynamics,horowitz2014thermodynamics,hartich_sensory_2016,barato_efficiency_2014,ito2013information,Bisker_2017,shiraishi_ito_sagawa_thermo_of_time_separation.2015}, i.e., aggregate
systems composed of two co-evolving smaller systems, whose states fluctuate according to 
independent noise processes (e.g., since they are physically separated and
so are connected to different parts of any shared thermodynamic reservoirs).  
However, given that many aggregate systems comprise more than just two interacting systems, more recently the
research community has started to consider fully multipartite processes (MPPs), with arbitrarily many constituent
systems~\cite{horowitz_multipartite_2015,ito2013information,wolpert_book_2018,wolpert.thermo.bayes.nets.2020,wolpert_kolchinsk_first_circuits_published.2020,wolpert_composite_systems_2021,wolpert2020minimal,loos2020irreversibility}.
(In the sequel I will refer to either to a set of ``systems'' that comprise an MPP, or to a set of ``subsystems''
that do so, as convenient.)

One of the most important properties of complex systems that can be modeled as MPPs is that we can often 
extend all previously derived, conventional TURs, which bound the EP of single systems in
terms of an associated current precision, to instead bound the 
EP of the full, aggregate, complex system in terms of current precision(s) of its 
constituent co-evolving systems. These new, extended TURs are often stronger than the conventional, single-system TURs
they are built from, which
did not consider the structure of the joint dynamics of the co-evolving systems in an MPP.
Furthermore, the new
TURs often apply, bounding the global EP of the full, complex system,
even if the full system does not meet any of the criteria for a conventional, single-system TUR to apply.
(All that is needed is that there is some subset of the constituent systems that meets such a set of criteria). 
Moreover, in some cases different conventional TURs apply to different constituent systems. These
new results show that in such cases, those distinct 
conventional TURs
can be combined, to lower-bound the EP of the full, complex system in terms of current precisions of the associated systems. 

More generally, in an MPP we can often ``mix and match'' completely different types of
bounds involving EP, not just TURs. For example, it may be that one set of systems in an MPP meet the conditions required
for the canonical TUR, another
(perhaps overlapping) set of systems meet the conditions required for an SLT, some third set of systems 
meet the conditions required for a first-passage time bound, but the full, complex system does not meet the conditions
for \textit{any} of those results to apply. So we cannot apply any one of those results to bound the
global EP generated by the complex system. However, by applying those results separately, to bound the
EP generated by the associated
subset of systems for which they \textit{do} apply, we can combine them to bound the global EP of the complex system.
%

In this chapter I illustrate how we can mix and match the bounds this way, thereby strengthening
the second law of thermodynamics. First, in \cref{sec:terminology}, I formally define MPPs and introduce ``units'',
which are sets of systems in an MPP that evolve according to their own, self-contained CTMC.
Then in \cref{sec:thermo_MPPs} I introduce the elementary definitions of  thermodynamic quantities in MPPs,
along with important associated constructions. Next, in \cref{sec:FTs} I derive the fluctuation theorems for FTs.
Finally, in \cref{sec:mix_match} I illustrate some of the ways to mix and match all these results to provide new kinds of lower bounds on global EP.

By definition, in an MPP each of the constituent systems evolves due to coupling with its own, independent set of thermodynamic reservoirs.
As a result it is impossible for two systems to change state exactly simultaneously in an MPP. In some situations
though, this restriction does not hold. Importantly, all of the results in this paper can be extended to such situations, in which
some thermodynamic reservoirs cause simultaneous fluctuations in more than one system at once; see~\cite{wolpert_composite_systems_2021}.
(That extension requires extra notation, which is why it is not pursued here.)

All proofs and formal definitions not in the text are in the appendices.


\section{Multipartite processes terminology}
\label{sec:terminology}

I write $\NN$ for a set of $N$ systems, with finite state spaces
$\{X_i : i = 1, \ldots N\}$. For any system $i$, I write $-i$ as shorthand to refer to the other $N-1$ systems,
and write $x_{-i}$ to refer to the vector of all values $x_j$ except for the value $j = i$.
 I write the joint (vector) state space of $\NN$ as $X$, with elements $x$. I also write
$\vX$ for the set of all trajectories $\vx$ of the joint system across some (often implicit) time interval.
I use $\delta(\cdot, \cdot)$ to indicate the Kronecker delta function.


Following~\cite{horowitz_multipartite_2015,wolpert2020minimal}, I assume that each system is
in contact with its own reservoir(s). This implies that
the probability is zero that any two systems change state simultaneously.
Therefore there is a set of time-varying stochastic rate matrices, 
$\{K^{x'}_x(i; t) : i = 1, \ldots, N\}$, where for all $i$, $K^{x'}_x(i; t) = 0$ if $x'_{-i} \ne x_{-i}$, and
the joint dynamics of the set of all the systems is given by~\cite{horowitz2014thermodynamics,horowitz_multipartite_2015}
\eq{
\frac{d p_x(t)}{dt} &= \sum_{x'} K^{x'}_{x}(t) p_{x'}(t)   =  \sum_{x'} \sum_{i \in \NN} K^{x'}_{x}(i; t) p_{x'}(t)
}
Such a scenario formally defines a ``multipartite process'' (MPP).
For any $A \subseteq \NN$ I define $
K^{x'}_{x}(A; t) := \sum_{i \in A} K^{x'}_{x}(i; t).$

\begin{figure}[tbp]
\includegraphics[width=75mm]{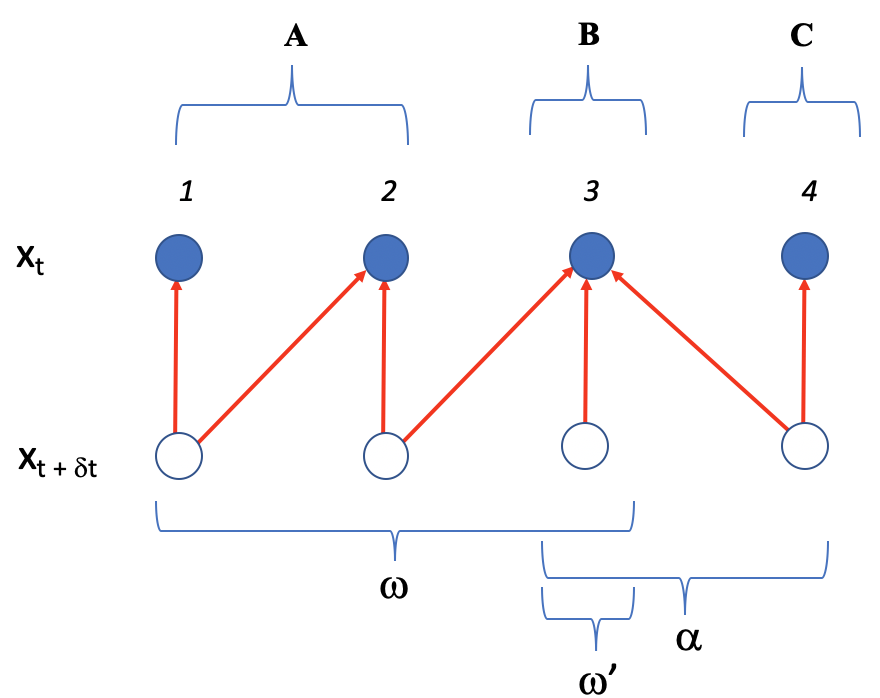}
\caption{Four systems, $\{1, 2, 3, 4\}$ interacting in a MPP.
The red arrows indicate dependencies in the associated four rate matrices. 
$B$ evolves independently, but is continually observed by $A$ and $C$. So the statistical coupling
between $A$ and $C$ could grow with time, even though their rate matrices do not involve one
another. Three examples of units are the sets $\oo, \oo', \alpha$
indicated at the bottom: $r(1) = r(2) = \oo$,
$r(4) = \alpha$, and $r(3) = \oo'$. 
}
\label{fig:1}
\end{figure}

As a concrete illustration, consider the scenario investigated in~\cite{hartich_sensory_2016,bo2015thermodynamic},
in which receptors in the wall of a cell sense the concentration of a ligand in the intercellular medium,
and those receptors are in turn observed by a ``memory'' subsystem inside the cell. Modify this scenario by introducing a second
cell, which is observing the same external
medium as the first cell. Assume that the cells are far enough apart physically so that
their dynamics are independent of one another. This gives us the precise scenario in \cref{fig:1}, where 
subsystem $3$ is concentration in the external medium, subsystem $2$ is the state of the receptors of 
the first cell, subsystem $1$ is the memory subsystem of the first cell,
and subsystem $4$ is the state of the receptors of the second cell. 

For each system $i$, I write $r(i; t)$ for any set of systems at time $t$ that includes $i$ such that
we can write
\eq{
K^{x'}_x(i; t) = K^{x'_{r(i;t)}}_{x_{r(i;t)}}(i; t) \delta(x'_{-r(i;t)}, x_{-r(i;t)})
\label{eq:def_unit_rate}
} 
for any stochastic rate matrix $K^{x'_{r(i;t)}}_{x_{r(i;t)}}(i; t)$ defined over $X_{r(i;t)}$.  Note that in general, for any given $i$, there
are multiple such sets $r(i;t)$.

A \textbf{unit} $\oo$ (at an implicit time $t$) is a set of systems such that $i \in \oo$ implies that $r(i;t) \subseteq \oo$.
Any intersection of two units is a unit, as is any union of two units. 
(See  \cref{fig:1} for an example.)
A set of units that covers $\NN$ and is closed under intersections is a \textbf{unit structure}, typically written as $\NN^*$.
In general, a given MPP can be described with more than one unit structure.
Except where stated otherwise, I focus on unit structures which do not include $\NN$ itself as a member. 
From now on I assume there are prefixed time-intervals in which $\NN^*$ doesn't change,
and restrict attention to such an interval. (This assumption holds in all of the papers mentioned above.)

For any unit $\oo$, I write $
K^{x'_\oo}_{x_\oo}(\oo; t) := \sum_{i \in \oo} K^{x'_{\oo}}_{x_{\oo}}(i; t)$.
So 
\eq{
K^{x'}_x(\oo; t) &:= \sum_{i \in \oo} K^{x'}_{x}(i; t) \\
	&= K^{x'_\oo}_{x_\oo}(\oo; t)  \delta(x'_{-\oo}, x_{-\oo})
}
Crucially, at any time $t$, for any unit $\omega$, $p_{x_{\omega}}(t)$ evolves as a self-contained CTMC with rate
matrix $K^{x'_{\omega}}_{x_\omega}(\omega; t)$:
\eq{
\frac{d p_{x_\oo}(t)}{dt} 
	&= \sum_{x'_\oo} K^{x'_\oo}_{x_\oo}(\oo; t) p_{x'_\oo}(t)
\label{eq:15aa}
}
(See App.\,A in~\cite{wolpert2020minimal} for proof.) Therefore any unit obeys all the usual stochastic thermodynamics
theorems, e.g., the second law, the FTs, the TURs, etc. In general,
this is not true for an arbitrary set of systems in an MPP~\cite{wolpert2020minimal}.

As an example,~\cite{hartich_sensory_2016,bo2015thermodynamic} considers an MPP where receptors in a cell wall observe the
concentration of a ligand in a surrounding medium, without any back-action from the receptor onto that concentration level~\cite{sagawa2012fluctuation,van2020thermodynamic,horowitz2011designing,shiraishi_ito_sagawa_thermo_of_time_separation.2015,wachtler2016stochastic,verley_work_2014}. In addition to the medium and 
receptors, there is a memory that observes the state of those receptors. We can extend that scenario, to include a second set of
receptors that observe the same medium (with all variables appropriately coarse-grained).  
\cref{fig:1} illustrates this MPP; system $3$ is the concentration level,
system $2$ is the first set of receptors observing that concentration level, system $1$ is the memory, 
and system $4$ is the second set of receptors.



\section{Thermodynamics of multipartite processes}
\label{sec:thermo_MPPs}

Taking $k_B = 1$, I write the inverse temperature of reservoir $k$ for system $i$ as $\beta_i^k$,
 the associated chemical potentials as $\mu_i^k$ (with  $\mu_i^k = 0$ if $k$ is a heat bath), and
any associated conserved quantities as $n_i^k(x_i)$.
Accordingly I write the rate matrix of system $i$ as
$K_x^{x'}(i; t) = \sum_k K_x^{x'}(i; k, t)$.
%
Any fluctuations of $x$ in which only $x_i$ changes  are determined by exchanges between $i$ and its reservoirs.
Moreover, since we have a MPP, the rate matrices $K(j;t)$ for $j \ne i$ 
must equal zero for such a fluctuation in the state of $i$. Therefore, writing  $H_{x}(t)$ for the
\textbf{global} Hamiltonian, thermodynamic consistency~\cite{rao2018conservation,van2013stochastic} says that for all $i, k, t, x, x'$,
\eq{
\!\!\!\! \!\!\!\! \!\!\!\! \ln \dfrac{K^{x'}_{x}(i; k, t)}{K^{x}_{x'}(i; k, t)} 
	= \beta_i^k\left(\left[H_{x'}(t) - H_{x}(t)\right] + \mu_{i}^k\left[n_i^k(x_i) - n_i^k(x'_i)\right]\right)
\label{eq:nldb1}
}
so long as $x_{-i} = x'_{-i}$.

The LHS of \cref{eq:nldb1} cannot depend on values $x_j$ or $x'_j$ for $j \not \in r(i)$. Since we are only
ever interested in differences in energy, this means that we must be able to rewrite
the RHS of \cref{eq:nldb1} as
\eq{
\beta_i^k\left(\left[H_{x'_{r(i)}}(i;t) - H_{x_{r(i)}}(i;t)\right] + \mu_{i}^k\left[n_i^k(x_i) - n_i^k(x'_i)\right]\right)
\label{eq:nldb}
}
for some \textbf{local} Hamiltonians $H_{x_{r(i)}}(i; t)$.  This condition is called
\textbf{subsystem} LDB (SLDB).\footnote{SLDB does not always hold, i.e., there are unit structures
that violate \cref{eq:nldb1}  and so are not thermodynamically consistent. (A fully worked-out example is
given  in \cref{app:two-way-observation}.)
Unless explicitly stated otherwise, I assume throughout this paper that we do not have such a unit
structure, and so SLDB does hold.
Some sufficient conditions for SLDB to hold exactly are derived in \cref{app:exact_ldb}, and
sufficient conditions for it to hold to high accuracy are derived in \cref{app:approx_ldb}.}

To define trajectory-level quantities, first, for any set of systems in $\alpha$,
define the \textbf{local} stochastic entropy 
as
\eq{
s^{\alpha}(\vx_\alpha(t)) &:= -\ln p_{\vx_\alpha(t)}(t) 
\label{eq:stoch_entropy}
}
In general I will use the prefix $\Delta$ to
indicate the change of a variable's value from the beginning of a process at $t = 0$ to its end, at $t = t_f$, e.g., 
\eq{
\Delta s^{\alpha}(\vx_\alpha) &:= \ln p_{\vx_\alpha(0)}(0) - \ln p_{\vx_\alpha(t_f)}(t_f)
\label{eq:43}
}
The trajectory-level entropy flows (EFs) between a system $i$ and its reservoirs
along a trajectory is written as $Q^i(\vx)$.
The \textbf{local} EF into $\alpha$ is the total EF into the systems $\alpha$ from its reservoirs during $[0,t_f]$:
\eq{
Q^\alpha(\vx) &:= \sum_{i \in \alpha} Q^i(\vx)
\label{eq:19}
}
(See \cref{app:def_EF_units} for a fully formal definition.)

The \textbf{local} EP of any set of systems $\alpha$ is 
\eq{
\sigma^\alpha(\vx) &:=  \Delta s^\alpha(\vx) - Q^\alpha(\vx)
\label{eq:local_EP}
}
This can be evaluated by combining \cref{eq:43} and the expansion of local EF in 
\cref{app:def_EF_units}.

For any unit $\oo$, the expectation of $\sigma^\oo$ is
non-negative.\footnote{This is not true for some quantities called ``entropy production'' in the literature,
nor for the analogous expression ``$\sigma_x$'' defined just before Eq.\,(21) in~\cite{shiraishi2015fluctuation}.}
In addition, due to the definition of a unit, the definition of $Q^i(\vx)$, and \cref{eq:19},
the EF into (the systems in) $\oo$ along trajectory $\vx$ is only a function of $\vx_\oo$.
So we can write $Q^\oo(\vx) = Q^\oo(\vx_\oo)$. 
Since $\Delta s^\oo(\vx)$ also only depends on $\vx_\oo$, this means that
we can write $\sigma^\oo(\vx)$ as $\sigma^\oo(\vx_\oo)$.

Setting $\alpha = \NN$ in \cref{eq:43,eq:local_EP} allows us to
define {global} versions of those trajectory-level thermodynamic quantities. In particular, the \textbf{global} EP is
\eq{
\sigma^\NN(\vx) &:=  \Delta s^\NN(\vx) - Q^\NN(\vx)
\label{eq:global_EP}
}
There are several ways to expand the RHS of \cref{eq:global_EP}.
One of these decompositions, discussed in \cref{app:chi_decomp},
involves extending what is called the 
``learning rate'' in the literature~\cite{barato_efficiency_2014,hartich_stochastic_2014,hartich_sensory_2016,matsumoto2018role,Brittain_2017}
in two ways: to apply to MPPs that have more than two systems, and to apply to MPPs even if they are not in
a stationary state.

Here  
I focus on a
different decomposition however.
%
To begin, let ${\NN^*} = \{\oo_j : j = 1, 2, \ldots, n\}$ be a unit structure.
Suppose we have a set of functions indexed by the units, $\{f^{\oo_j} : \vX \rightarrow \R\}$. The
associated \textbf{inclusion-exclusion sum} (or just ``in-ex sum'') is 
defined as
\eq{
\widehat{\sum_{\oo' \in {\NN^*}}} f^{\oo'}(\vx) &:= \sum_{j = 1}^n f^{\oo_j}(\vx) - \sum_{1 \le j < j' \le n} f^{\oo_j \cap \oo_{j'}}(\vx) \nonumber \\
	& \qquad+ 	\sum_{1 \le j < j' < j'' \le n} f^{\oo_j \cap \oo_{j'} \cap \oo_{j''}}(\vx) - \ldots
\label{eq:in_ex_gen}
}
The time-$t$ \textbf{in-ex information} is then defined as
\eq{
&\II^{\NN^*}(\vx(t)) := \left[\widehat{\sum_{\oo \in {\NN^*}}}   s^\oo(\vx(t))\right] - s(\vx(t)) 
\label{eq:in_ex_info}
}
(A related concept, not involving a unit structure, is called ``co-information'' in~\cite{bell2003co}.)

As an example, if ${\NN^*} $ consists of two units, 
$\oo_1, \oo_2$, with no intersection, then the expected in-ex information at time $t$ is just
the mutual information between those units at that time.
More generally, if there an arbitrary number of units in ${\NN^*}$
but none of them overlap, then the expected in-ex information is what is called the 
``multi-information'', or
``total correlation''~\cite{ting1962amount,wolpert2020minimal} in the literature, 
\eq{
I\left( X^{\oo_1}(t); X^{\oo_2}(t);  \ldots \right) &:=  \left(\sum_\oo \langle s(X^\oo(t)) \rangle\right)  - \langle s(X(t)) \rangle 
\label{eq:total_correlation}
}
where $\langle \cdot \rangle$ indicates an expectation under the indicated random variable.

Intuitively, total correlation tells us how much (Shannon) information there
is in the set of all the variables, beyond that given by each of the variables considered independently. 
It can be viewed as a generalization of mutual information, to the case of more than two random variables.

Since unit structures are closed under intersections, if we apply the inclusion-exclusion principle
to the expressions for local and global EF
we get
\eq{
Q(\vx) 	&= \widehat{\sum_{\oo \in {\NN^*}}}  {Q}^\oo(\vx)
\label{eq:heat_decomp}
} 
Combining this with \cref{eq:43,eq:local_EP,eq:global_EP,eq:in_ex_info} gives
\eq{
\sigma^\NN(\vx) &=   \widehat{\sum_{\oo \in {\NN^*}}}  \sigma^{\oo}(\vx) - \Delta \II^{{\NN^*}}(\vx)
\label{eq:global_EP_decomp_in_ex}
}
Taking expectations of both sides of \cref{eq:global_EP_decomp_in_ex} we get
\eq{
\langle \sigma^\NN \rangle &=   \widehat{\sum_{\oo \in {\NN^*}}} \langle \sigma^{\oo}\rangle -\langle \Delta \II^{{\NN^*}}\rangle
\label{eq:29}
}
(See App.\,E in~\cite{wolpert2020minimal}.)

As a simple example,
if there are no overlaps between any units,
then \cref{eq:29} reduces to
\eq{
\label{eq:in_ex_decomp0}
\langle \sigma^\NN \rangle &= \sum_\oo \langle \sigma^\oo \rangle -   \langle\Delta I\left( X^{\oo_1}; X^{\oo_2};  \ldots \right)  \rangle
		\\
	&\ge  -   \langle\Delta I\left( X^{\oo_1}; X^{\oo_2};  \ldots \right)  \rangle
\label{eq:in_ex_decomp}
}
(where the second line relies on the fact that each unit evolves according to its own rate matrix, and therefore
obeys the second law on its own). 
\cref{eq:in_ex_decomp} is a strengthened form of the second law of thermodynamics, reflecting the 
extra constraints on the dynamics of the full system given by  unit structure, first derived
in~\cite{wolpert_thermo_comp_review_2019}. It tells us that when the units do not overlap, the
EP is lower-bounded by the drop in total correlation among the units, which is non-negative and
typically is strictly positive. 

To illustrate this case where no units overlap, consider a process that erases three bits in parallel, where each
bit evolves as its own unit, and generates no local EP as it gets erased. Suppose that the initial joint distribution over the three bits assigned
probability $1/2$ to the joint state where all bits are up, and probability $1/2$ to the joint state where
all are down. Then \cref{eq:in_ex_decomp} tells us that even though each bit undergoes
a thermodynamically reversible process when considered on its own, and even though 
each bit's final state is independent of the states of the other bits, the
EP generated during the full process is $2 \ln 2$. In contrast, if the rate matrix allowed 
the dynamics of each bit to depend on the state of the other bits, then even though
that extra information had no effect on the final outcome, the minimal EP would only
be $0$. 

Finally, it is well-known that coarse-graining increases the EP of a system (due to the data-processing
inequality for KL divergence). In the current context that means that
\eq{
 \left\langle \sigma^\NN\right\rangle \ge \left\langle \sigma^\oo \right\rangle
\label{eq:38}
}
%
for any unit $\oo \subset \NN$. 
Combining \cref{eq:38} with \cref{eq:29} gives
\eq{
{\widehat{\sum}}_{\oo'\in {\NN^*}} \langle \sigma^{\oo'}\rangle - \langle\Delta \II^{{\NN^*}}\rangle \ge 
 \langle \sigma^{\oo}\rangle
\label{eq:23d}
}
(As illustrated below, this bound can help simplify certain calculations.)

In addition, suppose we have a set of units $\{\AAA_i\}$ which do not 
necessarily form a unit structure. Writing $\cup_i \AAA_i = \AAA$, since the union of units is a unit, \cref{eq:38} tells us that
\eq{
\left\langle \sigma^\NN \right\rangle \ge  \left\langle \sigma^{\AAA} \right\rangle
}
Therefore for any unit structure $\AAA^*$ which covers $\AAA$ and which also includes the units $\{\AAA_i\}$,
the global expected EP is lower-bounded by
\eq{
\left\langle \sigma^\NN \right\rangle \ge {\widehat{\sum}}_{\alpha \in \AAA^*} \langle \sigma^{\alpha}\rangle - \langle\Delta \II^{{\AAA^*}}\rangle
\label{eq:28}
}


\section{Multipartite process fluctuation theorems}
\label{sec:FTs}
Write $\tilde{\vx}$ for $\vx$ reversed in time, i.e., $\tilde{ \vx}(t) = \vx(t_f - t)$.
Write $\vP(\vx)$  for the probability density function over trajectories
generated by {starting} from the initial distribution $p_x(0)$; Also
write $\tilde{\vP}(\vx)$ for the probability density function over trajectories
generated by {starting} from the ending distribution $p_x(t_f)$, and then evolving 
according to the time-reversed sequence of rate matrices, $\tilde{K}(t) = K(t_f-t)$.
(Formal definitions of these density functions are given in 
\cref{app:def_EF_units,app:DFT_vector}.)

Let $\AAA = \{\AAA_i\}$ be any set of units (not necessarily a unit structure) which includes in particular the unit $\cup_i \AAA_i$. 
Abusing notation, write $Q^\AAA(\vx)$ for the total EF involving the systems in $\cup_i \AAA_i$ under trajectory $\vx$.
Similarly, write
$\sigma^{ \AAA}(\vx)$ for the total EP generated by all the systems in  $\cup_i \AAA_i$ under $\vP$.
(Note that since $\cup_i \AAA_i$ is a union of units, in fact we can write $\sigma^{ \AAA}(\vx)$ as $\sigma^{ \AAA}(\vx_\AAA)$.)

Next, define $\vec{\sigma}^\AAA$ as the vector whose
components are the local EP values $\sigma^\alpha$ for all $\alpha \in \AAA$.
The following detailed fluctuation theorem (DFT) concerning these vectors is derived in \cref{app:DFT_vector}:
\eq{
\ln \left[\dfrac{\vP(\vec{\sigma}^\AAA)}{ \tilde{\vP}(-\vec{\sigma}^\AAA)}\right]  &= \sigma^{ \AAA}
\label{eq:50}
}
where $ \tilde{\vP}(-\vec{\sigma}^\AAA)$
is the joint probability that the vector of EP values under ${\tilde {\vP}}$ is $-\vec{\sigma}^\AAA$.

It is important to note that \cref{eq:50} holds even though in general the different units
may overlap, so that some components of their trajectories are always identical. As a result
of such overlap, in general it is not true that  $\sigma^\AAA = \sum_{\alpha \in \AAA} \sigma^\alpha$
(See \cref{eq:global_EP_decomp_in_ex}.) In contrast,
the vector-valued DFT in Eq.\,47 of~\cite{garcia2012joint} (equivalently, Eq.\,2 of~\cite{garcia2010unifying})
only applies when the global EP is a sum of the local EPs, and so is a substantially
more restricted result than \cref{eq:50}. (See discussion in \cref{app:DFT_vector}.)
%

Subtracting instances of \cref{eq:50} evaluated for different choices of $\AAA$ 
gives conditional DFTs, which in turn give conditional integral fluctuation theorems (IFTs). 
As an example, subtract \cref{eq:50} for $\AAA$ set to some singleton $\{\oo\}$ from 
\cref{eq:50} for $\AAA = \{\NN\}$. Converting the resulting DFT into an IFT in the usual way gives
\eq{
\left\langle e^{\sigma^\oo - \sigma^\NN} \,\vert\,  \sigma^\oo \right\rangle &= 1
\label{eq:28c}
}
for any specific EP value $\sigma^\oo$ such that both $\vP(\sigma^\oo)$ and $\tilde{\vP}(-\sigma^\oo)$ are nonzero.
Applying Jensen's inequality to \cref{eq:28c} tells us that for all values $\sigma^\oo$ that can occur with nonzero probability,
\eq{
\left\langle \sigma^\NN \,\vert\, \sigma^\oo \right\rangle \ge \sigma^\oo
\label{eq:37}
}
\cref{eq:37} is a strictly stronger version of \cref{eq:38}, establishing that the inequality in \cref{eq:38} holds for
any single possible observed value of the local EP $\sigma^\oo$, not just for the average value. 

One can extend the definition of total correlation given in \cref{eq:total_correlation}, so that 
rather than tell us how much of the information in the joint state of the set of systems at a single moment $t$
is given by the statistical coupling among the states of the individual systems, it tells us much of the information in the 
the joint {trajectory} of
the set of systems is given by the statistical coupling among the {trajectories} of the individual systems.
Going further, one can normalize that measure by changing it to
tell us how much extra statistical coupling there is among the joint forward trajectories 
in comparison to the coupling among the associated backwards trajectories. We do this by replacing each of the entropies
of forward trajectories in the definition of total correlation with a relative entropy, between those forward
trajectories and the associated backward trajectories. This gives us the definition of the \textbf{multi-divergence} for a set of 
units $\AAA$~\cite{wolpert_thermo_comp_review_2019}:
\eq{
&  I\left(\vP(\vec{\sigma}^\AAA) \; ||\; \tilde{\vP}(-\vec{\sigma}^\AAA)\right) \nonumber \\
&\qquad :=		D\left(\vP(\vec{\sigma}^\AAA) \; ||\; \tilde{\vP}(-\vec{\sigma}^\AAA)\right) 
		- 	\sum_{\AAA_i \in \AAA} D\left(\vP({\sigma}^{\AAA_i}) \;||\; \tilde{\vP}(-\sigma^{\AAA_i})\right)
}
where $D(. \,||\, .)$ is relative entropy.

It is straightforward to confirm that
\eq{
\left\langle \sigma^\NN \right\rangle &\ge \sum_{i = 1}^m \left\langle \sigma^{\AAA_i}  \right\rangle + I\left(\vP(\vec{\sigma}^\AAA) \; ||\; 		
				\tilde{\vP}(-\vec{\sigma}^\AAA)\right) 
\label{eq:22}
}
Moreover, this bound becomes a strict equality if $\cup_i \AAA = \NN$,
even if the units overlap. (See \cref{app:other_implications}.)

Note that \cref{eq:22} involves a normal sum of local EPs, not an in-ex sum.
Since each term in that sum is non-negative, so is the full sum. This means
that  \cref{eq:22} provides a purely information-theoretic lower bound on EP which \textit{always} holds:
\eq{
\left\langle \sigma^\NN \right\rangle &\ge  I\left(\vP(\vec{\sigma}^\AAA) \; ||\; 		
				\tilde{\vP}(-\vec{\sigma}^\AAA)\right) 
\label{eq:30b}
}
There is no corresponding universal purely information-theoretic bound arising from \cref{eq:29}, because that inequality involves an in-ex sum
of local EPs, and that in-ex sum can be negative.

In addition, multi-divergence is guaranteed to be non-negative in many scenarios,
even if the units overlap, and even if the states of all the systems are changing in time. 
For example, multi-divergence is guaranteed to be non-negative so long as $\tilde{\vP}(-\vec{\sigma}^\AAA)$ is uniform
over its support. (This is due to the fact that total correlation is non-negative.)
As an illustration, $\tilde{\vP}(-\vec{\sigma}^\AAA)$ is uniform if the rate matrices are time-homogeneous, the Hamiltonian is uniform
and unchanging, and the system relaxes to a uniform distribution by time 
$t_f$. (See \cref{app:other_implications}.)

In such scenarios where multi-divergence is guaranteed to be non-negative, \cref{eq:22} implies
\eq{
\left\langle \sigma^\NN \right\rangle \ge \sum_{i = 1}^m \left\langle \sigma^{\AAA_i} \right\rangle
}
i.e., in such cases the total expected EP is lower-bounded by the sum of the local EPs.
Therefore in such scenarios we can use any applicable nonzero lower bounds on 
one or more of the local EPs, $\langle \sigma^{\AAA_i} \rangle$, to 
provide a strictly positive lower bound on the global EP. (An example of this is given in the next section.)


It is worth pointing out that none of the results above {require} that the times when each
system is allowed to change its state are random. Those results apply just as well to scenarios
where the systems take turns changing their states, like those analyzed in~\cite{ito2013information,wolpert2018thermo_circuits,wolpert.thermo.bayes.nets.2020}.



\section{Extended TURs, SLTs, and strengthened second law}
\label{sec:mix_match}

The simplified ligand-sensing example in \cref{fig:1}
can be used to illustrate some consequences of these results, where we choose a unit structure
with three units, $\{X^{AB}, X^B, X^{BC}\}$.

To begin, suppose $x_B$ is constant during the process.
Physically, this means that we suppose that since the medium is so large and well-mixed,
the ligand concentration in the medium is constant during the 
process (to within the precision of the coarse-grained binning of $X_B$). 
So there is no change in the entropy of $B$ during the process, and since there are no changes in $B$'s state, there
are no flows between $B$ and any reservoirs it has. So $ \left\langle \sigma^{B} \right\rangle = 0$. Moreover,
 $AB$ and $BC$ are both units, and so subject to the second law. Therefore \cref{eq:29} gives
\eq{
\label{eq:25a}
\left \langle \sigma^\NN \right \rangle &=  \left\langle \sigma^{AB} \right\rangle + \left\langle \sigma^{BC}\right\rangle 
  -  \Delta I(A; C \,|\, B)  \\
	&\ge -   \Delta I(A; C \,|\, B) 
\label{eq:25}
}
where 
$I\left(A; C \,|\, B\right)(t) = S(A \,|\, B) + S(C \,|\, B) - S(A, C \,|\, B)$ is the conditional mutual
information at time $t$ between $x_A$ and $x_C$, given $x_B$~\cite{cover_elements_2012}.
Although there are situations where $\Delta I(A; C \,|\, B)$ can be positive, it is straight-forward to prove 
that under our assumption that $x_B$ is constant, $\Delta I(A; C \,|\, B) \le 0$ (see \cref{app:conditional_independence_rule}). 
In addition, $ \left\langle \sigma^{AB} \right\rangle + \left\langle \sigma^{BC}\right\rangle  \ge 0$ always.
Accordingly, so long as the conditional mutual information between 
$A$ and $C$ changes during the process, the magnitude of that change provides 
a lower bound on global EP that is stronger than the conventional second law
(in the common situation where the ligand concentration in the medium is constant).
%

Alternatively, in this situation where $x_B$ does not change,
we can rewrite \cref{eq:25a} as the lower bound
\eq{
\left \langle \sigma^\NN \right \rangle &\ge  \left\langle \sigma^{AB} \right\rangle + \left\langle \sigma^{BC}\right\rangle 
}
We can now plug in any of the bounds involving EP discussed in the introduction to lower bound the two terms on the
RHS, so long as those bounds do not violate the assumption that $x_B$ stays 
constant.\footnote{Note that there are many reasons why $x_B$ may be constant, in
addition to having the underlying system $B$ be very large, as in the ligand-sensing example. For example, $x_B$ will stay
constant at its initial value whenever the energy barriers between states of $B$ are very high compared to the
size of the thermal fluctuations caused by coupling to $B$ reservoirs, if the rate matrices of the system never change.}
For example, in~\cite{shiraishi2019information}, it is shown that the EP generated during $[0, t_f]$ by any system thermally relaxing from
its initial distribution is bounded below by $D(p(0) \;||\; p(t_f))$, the relative entropy between the initial and final distributions
of states of the system.
Similarly, in the original SLT paper,~\cite{shiraishi_speed_2018}, it is shown that for any system evolving under
a CTMC from $p(0)$ to $p(t_f)$,  the total variation distance
between $p(0)$ and $p(t_f)$, $L(p(0), p(t_f))$, is bounded by
\eq{
\dfrac{L(p(0), p(t_f))^2}{2 t_f \langle A\rangle} \le \langle\sigma \rangle 
}
where $ \langle A\rangle$ is the time-integrated expected ``activity'' of the system during the interval $[0, t_f]$. 
Combining gives 
\eq{
\left \langle \sigma^\NN \right \rangle &\ge \dfrac{L(p_{AB}(0), p_{AB}(t_f))^2}{2 t_f \langle A_{AB}\rangle} + D(p_{BC}(0) \;||\; p_{BC}(t_f)) 
}
This illustrates how we can ``mix and match'' lower bounds on EP that apply to subsets of the full set of systems,
in order to derive a lower bound on the global EP of the full set of systems.

Next, evaluate \cref{eq:23d} for $\oo = AB$, to get
\eq{
& \left\langle \sigma^{AB} \right \rangle + \left\langle \sigma^{B} \right \rangle + \left\langle \sigma^{BC} \right \rangle
	- 3  \left\langle \sigma^{B} \right \rangle  +  \left\langle \sigma^{B} \right \rangle \nonumber \\
&\qquad \ge  \Delta S(X^{AB}) + \Delta S(X^{B}) + \Delta S(X^{BC}) - 3 \Delta S(X^{B}) + \Delta S(X^{B}) \nonumber \\
&\qquad\qquad\qquad - \Delta S(X^{ABC}) + \left\langle \sigma^{AB} \right \rangle
}
This reduces to
\eq{
\left\langle \sigma^{BC} \right \rangle - \left\langle \sigma^B \right\rangle \ge   \Delta I(A; C \,|\, B)
\label{eq:23dd}
}
Now instead of assuming system $B$ is unchanging in time, as above, assume it is subject to a TUR.
For example, it might be in an NESS, so that we can apply the canonical TUR (recall
\cref{eq:ness_tur}), which here takes the form
\eq{
\langle \sigma^B \rangle \ge \dfrac{2\langle J_B\rangle^2 }{{\mbox{Var}}(J_B)} 
\label{eq:36}
}
where $J_B(\vx_B)$ is the net value during $[0, t_f]$ of an (arbitrary) current measuring transitions among $B$'s
states.
Plugging \cref{eq:36} into \cref{eq:23dd}, applying \cref{eq:38} for $\oo = BC$, and then applying \cref{eq:38} 
for $\oo = B$, we establish that
\eq{
\langle \sigma^\NN\rangle \ge \dfrac{2\langle J_B\rangle^2 }{{\mbox{Var}}(J_B)}  + \max\left[ \Delta I(A; C \,|\, B), 0\right] 
\label{eq:27aa}
}
\cref{eq:27aa} can be viewed as a 
new kind of TUR, bounding global EP in terms of current precision plus a purely information-theoretic term.
In this sense, it extends the canonical TUR similarly to how the second law can be extended
to account for feedback control, by adding a purely information-theoretic term to the second law's lower bound on EP~\cite{parrondo2015thermodynamics,sagawa2012fluctuation}.

Importantly, if the full system is not at an NESS, then
we cannot apply the canonical TUR directly, to bound the global EP in terms of
an arbitrary, system-wide current $J_{ABC}$. Nonetheless, \cref{eq:27aa} means 
we can use the canonical TUR to bound the global EP by applying it to
just the (local) system $B$.  
%

This reasoning can be generalized beyond the canonical TUR; \cref{eq:23dd} can be used to derive inequalities like \cref{eq:27aa} for \textit{any}
previously derived TUR like those discussed in the introduction. 
These new inequalities will bound global EP so long as system $B$ obeys the condition for 
that previously derived TUR. This is true even
if the full system does not obey that TUR, and / or we are not able to measure currents other than those in $B$.
Indeed, this kind of reasoning can be applied not just using TURs, but using any of the conventional stochastic thermodynamic
theorems in which EP provides an upper bound on other quantities of interest, e.g., SLTs, first-passage bounds, 
stopping condition bounds, etc.

%
%

Next, choose $\AAA = \{AB, BC\}$ in  \cref{eq:22} to get
\eq{
\left\langle \sigma^\NN \right\rangle \ge \left\langle \sigma^{AB}  \right\rangle + \left\langle \sigma^{BC}  \right\rangle 
	+ I\left(\vP(\sigma^{AB}, \sigma^{BC}) \; ||\; \tilde{\vP}(-\sigma^{AB}, -\sigma^{BC})\right)
\label{eq:28aa}
}
Suppose the rate matrices are time-homogeneous, the Hamiltonian is uniform
and unchanging, and  the full system $ABC$ relaxes to a uniform distribution fixed point by time $t_f$.
As described in \cref{app:other_implications}, these three conditions imply
that $I\left(\vP(\sigma^{AB}, \sigma^{BC}) \; ||\; \tilde{\vP}(-\sigma^{AB}, -\sigma^{BC})\right) \ge 0$.
In such a situation, \cref{eq:28aa} gives another lower bound on global EP that is stronger than the second law,
\eq{
\left\langle \sigma^\NN \right\rangle \ge \left\langle \sigma^{AB}  \right\rangle + \left\langle \sigma^{BC}  \right\rangle 
}

In addition, whenever the Hamiltonian is unchanging in time, the relaxing unit $BC$ obeys the conditions 
for the arbitrary initial state TUR~\cite{liu2020thermodynamic}, as does the unit $AB$.
Therefore so long as  $I\left(\vP(\sigma^{AB}, \sigma^{BC}) \; ||\; \tilde{\vP}(-\sigma^{AB}, -\sigma^{BC})\right) \ge 0$,
\cref{eq:28aa} lower-bounds global EP in such cases in terms of a pair of current ``precisions'' in {different} units:
\eq{
\langle \sigma^\NN\rangle  &\ge  
			\dfrac{\langle t_f j_{AB}(t_f)\rangle^2}{ {\mbox{Var}}(J_{AB})} +
		\dfrac{\langle t_f j_{BC}(t_f)\rangle^2}{ {\mbox{Var}}(J_{BC})}
\label{eq:29aa}
}
($j_{BC}(t_f)$ is the derivative of the expectation of an arbitrary current in joint system $BC$, evaluated at $t_f$, and
$j_{AB}(t_f)$ is defined analogously; see~\cite{liu2020thermodynamic}.) 

Finally, combining \cref{eq:29,eq:28aa}
gives 
\eq{
 \left\langle \sigma^B \right\rangle \le  -\Delta I(A; C \,|\, B)
	- I\left(\vP(\sigma^{AB}, \sigma^{BC}) \; ||\; \tilde{\vP}(-\sigma^{AB}, -\sigma^{BC})\right)
\label{eq:30aa}
}
\cref{eq:30aa} shows that it is impossible to have both $\Delta I(A; C \,|\, B) > 0$ and $I\left(\vP(\sigma^{AB}, \sigma^{BC}) \; ||\; \tilde{\vP}(-\sigma^{AB}, -\sigma^{BC})\right) > 0$. Moreover,
suppose we know that $I\left(\vP(\sigma^{AB}, \sigma^{BC}) \; ||\; \tilde{\vP}(-\sigma^{AB}, -\sigma^{BC})\right) \ge 0$,
e.g., if $AB$ is in a NESS.
In this case
\cref{eq:30aa} \textit{upper}-bounds EP in the inter-cellular medium in terms of a purely
information-theoretic quantity, $-\Delta I(A; C \,|\, B)$, which also involves the cell wall receptor systems $A$ and $C$.
In fact, in this specific case that $AB$ is in an NESS, $B$ must also be in a NESS. So by the canonical NESS-based TUR,
\eq{
2 \dfrac{\langle J_{AB}\rangle^2}{ {\mbox{Var}}(J_{AB})} \le  -\Delta I(A; C \,|\, B)
}
This is a new kind of bound, establishing that if the conditional mutual information between $A$ and $C$ drops
significantly, then the precision of the current $J_B$ cannot be large.

\section*{Acknowledgments}

I would like to acknowledge the Santa Fe Institute for support.

$ $

\section*{APPENDIX A:  Entropy flow into systems and units}
\label{app:def_EF_units}

To fully define the entropy flow into a set of systems in an MPP we need to introduce some more notation.
Let $M(\vx)$ be the total number of state transitions during the time interval $[0,t_f]$ by all systems 
(which might equal $0$). If $M(\vx) \ge 1$, define $\eta_\vx : \{1, \ldots, M(\vx)\} \rightarrow \NN$ as the function that maps
any integer $j \in  \{1, \ldots, M(\vx)\}$ to the system that changes its state in the $j$'th transition.  Let $k(j)$ be the
associated function specifying which reservoir is involved in that $j$'th transition. (So for all $j$, $k(j)$ specifies
a reservoir of system $\eta(j)$.) Similarly,
let $\tau_\vx : \{0, \ldots, M(\vx)\} \rightarrow \NN$ be the function that maps
any integer $j \in  \{1, \ldots, M(\vx)\}$ to the time of the $j$'th transition, and maps $0$ to the time $0$. 

From now on, I leave the subscript $\vx$ on the maps $\eta_\vx$ and $\tau_\vx$ implicit.
So for example, $\eta^{-1}(i)$ is the set (of indices specifying) all state transitions at which system $i$ changes state in the trajectory $\vx$. 
More generally, for any set of systems $\alpha$, $\eta^{-1}(\alpha) := \cup_{i \in \alpha} \eta^{-1}(i)$ is the set
of  all state transitions at which a system $i \in \alpha$ changes state in the trajectory $\vx$. 

Given these definitions, the total entropy flow into system $i$ from its reservoirs during $[0,t_f]$ is defined as~\cite{van2015ensemble}
\eq{
Q^i(\vx) &:=  \sum_{j \in \eta^{-1}(i)}  
\beta_i^{k(j)}\bigg(H_{{\vx}_{r(i)}(\tau(j))}(i;\tau(j)) - H_{{\vx}_{r(i)}(\tau(j-1))}(i;\tau(j))  \nonumber \\
	&\qquad + \mu_{i}^{k(j)}\left[n_i^{k(j)}({\vx}_i(\tau(j-1)) - n_i^{k(j)}({\vx}_i(\tau(j))\right]\bigg) 
\label{eq:19a}
}
where I interpret the sum on the RHS to be zero if system $i$ never undergoes a state transition in trajectory $\vx$.

As mentioned in the text, the local EF into a unit $\alpha$ for trajectory $\vx$ is
just the sum of the EFs into all the systems in $\alpha$ for that trajectory.
%
%
Expanding, under \cref{eq:nldb},
\eq{
Q^\alpha(\vx) &= \sum_{i \in \alpha}	\sum_{j \in \eta^{-1}(i)}  
						\ln	\left[\dfrac{K^{{\vx}_{r(i)}(\tau(j))}_{{\vx}_{r(i)}(\tau(j-1))}(i; k(j), \tau(j))}  {K^{{\vx}_{r(i)}(\tau(j-1))}_{{\vx}_{r(i)}(\tau(j))}(i; k(j), \tau(j))}\right]
\label{eq:local_EF}
}
So in the special case that $\alpha$ is a unit, 
\eq{
Q^\alpha(\vx) &= \sum_{i \in \alpha}	\sum_{j \in \eta^{-1}(i)}  
						\ln	\left[\dfrac{K^{{\vx}_{\alpha}(\tau(j))}_{{\vx}_{\alpha}(\tau(j-1))}(i; k(j), \tau(j))}  {K^{{\vx}_{\alpha}(\tau(j-1))}_{{\vx}_{\alpha}(\tau(j))}(i; k(j), \tau(j))}\right]
\label{eq:local_EF_unit}
}

Note that ${\vx}_{-i}(\tau(j-1)) = {\vx}_{-i}(\tau(j))$ for all systems $i$, for all $j \in \eta^{-1}(i)$, 
since the process is multipartite. Therefore the global EF can be written as
\eq{
Q(\vx) = \sum_i Q^i(\vx) 
	= \sum_{j =1}^{M(\vx)}  
						\ln	\left[\dfrac{K^{{\vx}(\tau(j))}_{{\vx}(\tau(j-1))}(\tau(j))}  {K^{{\vx}(\tau(j-1))}_{{\vx}(\tau(j))}(\tau(j))}\right]
\label{eq:30}
}

As a final, technical point, note that $\vx$ is fully specified by $M, \tau, k, \eta$ and the finite
list of the precise state transitions at the times listed in $\tau$ by the associated systems listed in $\eta$
mediated by the associated reservoirs listed in $k$. So the probability measure of a trajectory $\vx$ is 
\eq{
\vP(\vx) = P(M) P(\vx \,|\, M, \tau, k, \eta) P(\tau, k, \eta \,|\, M)
}
For each integer $M$, all the terms on the RHS are either probability distributions or probability density functions,
and therefore we can define the integral over $M, \tau, k, \eta, \vx$.
So in particular, 
$\delta$ function over trajectories in the equations in the main text is shorthand for a 
function that equals zero everywhere that its argument is nonzero, and such that its
integral equals $1$.


$ $

\section*{APPENDIX B: Example where global Hamiltonian cannot equal sum of local Hamiltonians under SLDB}
\label{app:two-way-observation}

The global Hamiltonian need not be a sum of the local Hamiltonians.
Indeed, it's possible that all local Hamiltonians equal one another and also equal the global Hamiltonian.
To see this, 
consider a scenario where there are exactly two distinct systems $i, j$, which both observe one another very closely, so that the rate matrices
of both of them have approximately as much dependence on the state of the \textit{other} system as on their own state.
In this scenario, there is a single unit, $r(i) = r(j) = \{i, j\} = \NN$. Assume as well that each
system is only connected to a single heat bath, with no other reservoirs.

If we took the global Hamiltonian to be
$H_x(t) = H_x(i; t) + H_x(j; t)$, its change under the fluctuation $(x'_i, x_{j}) \rightarrow (x_i, x_{j})$ would be
\eq{
H_{x_i, x_{j}}(i;t) - H_{x'_i, x_{j}}(i;t) + H_{x_i, x_{j}}(j;t) - H_{x'_i, x_{j}}(j;t)
\label{eq:d6}
}
This is the change in energy of the total system during the transition. Moreover, since this state transition 
only changes the state of system $i$ and since we assume the dynamics is a MPP, only the heat bath of system $i$ could 
have changed its energy during the transition. Therefore by conservation of energy,
the change in the energy of the bath of system $i$ must equal the expression
in \cref{eq:d6}. 

Under SLDB, that change in the energy
of the heat bath of system $i$ would be the change in the local Hamiltonian
of system $i$. 
Therefore to have SLDB be even approximately true, it would have to be the case that
\eq{
|H_{x_i, x_{j}}(i;t) - H_{x'_i, x_{j}}(i;t)|  \;&\gg\;   |H_{x_i, x_{j}}(j;t) - H_{x'_i, x_{j}}(j;t)|
\label{eq:d7}
}
for all  pairs $x, x'$.
%
Similarly, to address the case where a fluctuation to $x_j$ arises due to $j$'s interaction
with \textit{its} heat bath, we would need to have
\eq{
|H_{x_i, x_{j}}(j;t) - H_{x_i, x'_{j}}(j;t)| \;&\gg\;  |H_{x_j, x_{j}}(i;t) - H_{x_i, x'_{j}}(i;t)|
\label{eq:d8}
}
for all pairs $x, x'$.

Next, we need to formalize our requirement that ``the rate matrices
of both \{systems\} have approximately as much dependence on the state of the {other} system as on their own state''.
One way to do that is to require that the dependence on $x_j$ of (the ratio of forward and backward terms in) $i$'s rate matrix is 
not much smaller than the typical terms in that rate matrix, i.e., for all pairs $x, x'$,
\eq{
&\left| \left(H_{x_i, x_{j}}(i;t) - H_{x'_i, x_{j}}(i;t)\right)  - \left(H_{x_i, x'_{j}}(i; t) - H_{x'_i, x'_{j}}(i; t)\right) \right| \nonumber \\
	&\qquad \sim \;
\dfrac{\left| H_{x_i, x_{j}}(i;t) - H_{x'_i, x_{j}}(i;t)\right|  + \left| H_{x_i, x'_{j}}(i; t) - H_{x'_i, x'_{j}}(i; t) \right| }{2}  \nonumber \\
	&\qquad := \Delta_{i;x,x'}
\label{eq:d9}
}
Similarly, since system $j$ is also observing $i$,
\eq{
&\left| \left(H_{x_i, x_{j}}(j;t) - H_{x_i, x'_{j}}(j;t)\right)  - \left(H_{x'_i, x_{j}}(j; t) - H_{x'_i, x'_{j}}(j; t)\right) \right| \nonumber \\
	&\qquad \sim \;
\dfrac{ \left| \left(H_{x_i, x_{j}}(j;t) - H_{x_i, x'_{j}}(j;t)\right)  + \left(H_{x'_i, x_{j}}(j; t) - H_{x'_i, x'_{j}}(j; t)\right) \right| }{2}   \nonumber \\
	&\qquad := \Delta_{j;x,x'}
\label{eq:d10}
}

Plugging \cref{eq:d7} into the definition of $\Delta_{i; x, x'}$ and then using the fact that for any real numbers $a, b$, $|a - b| < |a| +|b|$, we get
\eq{
&\left| \left(H_{x_i, x_{j}}(j;t) - H_{x'_i, x_{j}}(j;t)\right)  - \left(H_{x_i, x'_{j}}(j; t) - H_{x'_i, x'_{j}}(j; t)\right) \right| \nonumber \\
	&\qquad \ll \Delta_{i;x,x'}
\label{eq:d11}
}
In addition, shuffling terms in the LHS of \cref{eq:d9} gives
\eq{
&\left| \left(H_{x_i, x_{j}}(i;t) - H_{x_i, x'_{j}}(i; t) \right)  - \left(H_{x'_i, x_{j}}(i;t)  - H_{x'_i, x'_{j}}(i; t)\right) \right| \nonumber \\
	&\qquad \sim \;  \Delta_{i; x,x'}
\label{eq:d12}
}

The analogous reasoning with \cref{eq:d8,eq:d10} gives
\eq{
&\left| \left(H_{x_i, x'_{j}}(i;t) - H_{x_i, x_{j}}(i;t)\right)  - \left(H_{x'_i, x'_{j}}(i; t) - H_{x'_i, x_{j}}(i; t)\right) \right| \nonumber \\
	&\qquad \ll \Delta_{j;x,x'}
\label{eq:d13}
}
and
\eq{
&\left| \left(H_{x_i, x_{j}}(j;t) - H_{x'_i, x_{j}}(j; t) \right)  - \left(H_{x_i, x'_{j}}(j;t)  - H_{x'_i, x'_{j}}(j; t)\right) \right| \nonumber \\
	&\qquad \sim \;  \Delta_{j; x, x'}
\label{eq:d14}
}

Comparing \cref{eq:d12,eq:d13} establishes that $\Delta_{i; x, x'} \ll \Delta_{j; x, x'}$. However,
comparing \cref{eq:d11,eq:d14} establishes that $\Delta_{j; x, x'} \ll \Delta_{i; x, x'}$. This contradiction shows
that it is not possible for the global Hamiltonian to be a sum of the two local Hamiltonians, given that
both systems ``observe one another very closely''. (See \cref{app:exact_ldb}.)

$ $

\section*{APPENDIX C: Alternative expansion of trajectory-level global EP}
\label{app:chi_decomp}

\subsection*{Definition of $\chi^\oo(\vx)$}
First, for any unit $\oo$, define 
\eq{
\chi^\oo(\vx) := \sigma^\NN(\vx) - \sigma^\oo(\vx)
\label{eq:chi_def}
}
In this subsection I show that 
\eq{
 \chi^\oo(\vx) 	&= \Delta s^{-\oo \,|\, \oo}(\vx) - Q^{-\oo}(\vx)
\label{eq:global_EP_decomp_windowing}
}
where
\eq{
\Delta s^{-\oo \,|\, \oo}(\vx) &= \Delta s^{-\oo, \oo}(\vx) - \Delta s^{\oo}(\vx) \\
	&= -\Delta \ln p_{\oo,-\oo}(\vx) +  \Delta \ln p_{\oo}(\vx) \\
	&=  -\Delta \ln p_{\NN}(\vx) +  \Delta \ln p_{\oo}(\vx)
}
is the change in the value of the conditional stochastic entropy of the entire system given the state of $\oo$.
Note that in general $-\oo$  will not be a unit. So the
entropy flow into the associated reservoirs, $Q^{-\oo}(\vx)$, may depend on the trajectory of systems outside of $-\oo$,
i.e., it may depend on $\vx_{\oo}$. 

As shorthand, from now on I leave the function $\tau$ implicit, so that for example, $\vx(\tau(j))$ gets shortened
to $\vx(j)$. (Note though that with slight abuse of notation, I still take $\vx(t_f)$ to mean the state of the
system at $t = t_f$ under trajectory $\vx$.)
Given any unit $\oo$, we can expand the global EP as



\eq{
\sigma^\NN(\vx) &=   \ln p_{\vx(0)}(0) - \ln p_{\vx(t_f)}(t_f) + \sum_{j =1}^{M(\vx)}  
						\ln	\left[\dfrac{K^{{\vx}(j-1)}_{{\vx}(j)}(\eta(j);j)}  {K^{{\vx}(j)}_{{\vx}(j-1)}(\eta(j);j)}\right]  
 																		\nonumber \\
	&=  \ln p^{X_\oo}_{\vx_\oo(0)}(0)) - \ln p^{X_\oo}_{\vx_\oo(t_f)}(t_f) + \sum_{j \in \eta^{-1}(\oo)} 
		\ln \left[\dfrac{K^{{\vx_\oo}(j-1)}_{{\vx_\oo}(j)}(\eta(j);j)} 
				 {K^{{\vx}_\oo(j)}_{{\vx_\oo}(j-1)}(\eta(j);j) }\right]   \nonumber \\
	&	\qquad\qquad+  \ln p^{X|X_\oo}_{\vx(0)}(0)  - \ln p^{X|X_\oo}_{\vx(t_f)}(t_f) 
				  + \sum_{j \in \eta^{-1}(-\oo)} 	\ln \left[\dfrac{K^{{\vx}(j-1)}_{{\vx}(j)}(\eta(j);j)}  				
				{K^{{\vx}(j)}_{{\vx}(j-1)}(\eta(j);j) }\right] 			\nonumber \\
	&= \sigma^\oo(\vx) + \ln p^{X|X_\oo}_{\vx(0)}(0)  - \ln p^{X|X_\oo}_{\vx(t_f)}(t_f)
				  + \sum_{j \in \eta^{-1}(-\oo)} 	\ln \left[\dfrac{K^{{\vx}(j-1)}_{{\vx}(j)}(\eta(j);j)}  				
				{K^{{\vx}(j)}_{{\vx}(j-1)}(\eta(j);j) }\right] 
\label{eq:27b}
}

Expand
\eq{
\Delta s^{{-\oo} |\oo}(\vx) &= \Delta s^{(-\oo,\oo) \,|\, \oo}(\vx)  \\
	&= \Delta s^{\NN \,|\, \oo}(\vx)  \\
		&=  \ln p^{\NN|\oo}_{\vx(0)}(0)  - \ln p^{\NN|\oo}_{\vx(t_f)}(t_f) 
}
This allows us to rewrite \cref{eq:27b} more succinctly as
\eq{
\chi^\oo(\vx) &=  \Delta s^{\NN \,|\, \oo}(\vx) 
		 + \sum_{j \in \eta^{-1}(-\oo)}	\ln \left[\dfrac{K^{{\vx}(j-1)}_{{\vx}(j)}(\eta(j);j)}  				
				{K^{{\vx}(j)}_{{\vx}(j-1)}(\eta(j);j) }\right] \nonumber \\
	&= \Delta s^{\NN \,|\, \oo}(\vx) - Q^{-\oo}(\vx)
\label{eq:27a}
}
which establishes \cref{eq:global_EP_decomp_windowing}.

\subsection*{FTs involving $\chi^\oo(\vx)$}

To gain insight into \cref{eq:27a}, define the counterfactual rate matrix $\underline{K}(t) := K(-\oo; t)$,
and let $\underline{\vP}$ be what the density over trajectories $\vx$
would have been if the system had evolved from the initial distribution $p_x(0)$ under $\underline{K}(t)$ rather than $K(t)$. 
Define $\Delta s^{{-\oo}}_{\underline{\vP}}(\vx)$ and $\sigma_{\underline{\vP}}(\vx)$ accordingly.
Then we can expand the second term on the RHS of \cref{eq:27a} as
\eq{
Q^{-\oo}(\vx) &= \sigma_{\underline{\vP}}(\vx) - \Delta s^{{-\oo}}_{\underline{\vP}}(\vx)
\label{eq:33maintext}
}
So the heat flow from the baths connected to $-\oo$ into the associated systems
is the difference between a (counterfactual) global EP
and a (counterfactual) change in the entropy of those systems.

We can iterate these results, to get more refined decompositions of global EP. For example,
let $\underline{\NN}^*$ be a unit structure of $\underline{K}$, the counterfactual
rate matrix defined just before \cref{eq:33maintext}.
Let $\oo$ be a unit in ${\NN^*}$ while
$\alpha$ is a unit in $\underline{\NN}^*$. Then we can insert \cref{eq:33maintext} 
into \cref{eq:global_EP_decomp_windowing} and apply that  \cref{eq:global_EP_decomp_windowing}  again,
to the resulting term $\sigma_{\underline{\vP}}$, to get
\eq{
\sigma^\NN(\vx)  &= \sigma^\oo(\vx)  + \left[\sigma^\alpha_{\underline{\vP}}(\vx) + \chi^{\alpha}_{\underline{\vP}}(\vx)
			+ \Delta s^{\NN \,|\, \oo}_\vP(\vx) - \Delta s^{{-\oo}}_{\underline{\vP}}(\vx)
\right]
\label{eq:c12}
}

Note that in general, $\alpha$ might contain systems outside of $\NN\setminus \oo$. As a result,
it need not be a unit of the full rate matrix $K$.
In addition, both (counterfactual) rates
$d \left\langle \sigma^\alpha_{\underline{\vP}} \right\rangle_{\underline{\vP}} /dt$
and  $d \left\langle \chi^\alpha_{\underline{\vP}} \right\rangle_{\underline{\vP}} /dt$ are non-negative.
However, if we evaluate those two rates under the actual density $\vP$ rather than
the counterfactual $\underline{\vP}$, it may be that one or the other of them is negative. This is just like how the expected values of
the analogous ``EP'' terms in~\cite{ito2013information,shiraishi2015fluctuation}, which concern a single
system, may have negative derivatives.

The trajectory-level decomposition of global EP in terms of $\chi^\oo$ can be exploited to get
other FTs in addition to those presented in the main text. For example,
if we plug \cref{eq:chi_def} into \cref{eq:28} instead of plugging in \cref{eq:global_EP_decomp_in_ex},
and then use  \cref{eq:global_EP_decomp_windowing},  we get
\eq{
\left\langle e^{- \chi^\oo}  \,|\, \sigma^\oo \right\rangle &= 
 \left\langle \exp \left( Q^{-\oo} - \Delta s^{\NN|\oo}\right) \;\bigg| \; \sigma^\oo \right\rangle  = 1 
\label{eq:32c}
}
In contrast, the analogous expression using the in-ex sum expansion in the main text is
\eq{
\left\langle e^{\left(\Delta \II^{{\NN^*}} + \sigma^\oo - {\widehat{\sum}}_{\oo'\in {\NN^*}} \sigma^{\oo'} \right)}
			 \;\bigg| \; \sigma^\oo  \right\rangle = 1   
\label{eq:32b}
}
Applying Jensen's inequality to \cref{eq:32b} allows us to bound the change in the conditional entropy
of all systems \textit{outside} of any unit $\oo$, given the joint state of the systems
within $\oo$, by the EF of those systems:
\eq{
\Delta S^{\NN \;|\; \oo} \ge Q^{\NN \setminus \oo}
} 
(\cref{eq:32c} can often be refined by mixing and matching among alternative expansions of $\chi^\oo$,
along the lines of \cref{eq:c12}.)

Applying Jensen's inequality to \cref{eq:32c}  shows that for all $\sigma^\oo$ with nonzero probability,
\eq{
\left\langle \Delta s^{\NN|\oo} - Q^{-\oo} \,\bigg|\, \sigma^\oo \right\rangle &\ge 0  
\label{eq:39aa}
}
Averaging both sides of \cref{eq:39aa} over all $\sigma^\oo$ gives
\eq{
 \Delta S^{\NN|\oo} &\ge  \left\langle Q^{-\oo} \right\rangle
\label{eq:40}
}
So  the expected heat flow into the systems outside of $\oo$ during any interval is upper-bounded
by the change during that interval in the value of the conditional entropy of the full system given the entropy of the unit $\oo$.

Since EPs, $\chi$'s, etc., all go to $0$ as $t_f \rightarrow 0$,
these bounds can all be translated into bounds concerning time derivatives. For example, in~\cite{wolpert2020minimal} it is
shown that $d \left\langle \chi^\oo \right\rangle /dt \ge 0$. Applying Jensen's inequality
to \cref{eq:32c} gives a strengthened version of this result: for any value
of $\sigma^\oo$ that has nonzero probability throughout an interval $t \in [0, t'>0]$, 
$d \left\langle \chi^\oo \,|\, \sigma^\oo \right\rangle / dt  \ge 0$ at $t=0$. 

%

\subsection*{Rate of change of expected $\chi^\oo(\vx)$}

I now show that $d \left\langle \chi^\oo \right \rangle /dt$ is  
the sum of two terms. The first term is the expected global EP rate under a counterfactual rate matrix.
The second term is (negative of) the derivative of the mutual information between $X_\oo$ and $X_{-\oo}$, under a counterfactual
rate matrix in which $x_{-\oo}$ never changes its state. 
(This second term is an extension of what is sometimes called the ``learning rate'' in~\cite{wolpert2020minimal,barato_efficiency_2014,hartich_stochastic_2014,hartich_sensory_2016,matsumoto2018role,Brittain_2017}
and is related to what is called ``information flow'' in~\cite{horowitz2014thermodynamics}.)
Both of these terms are non-negative. Plugging into \cref{eq:chi_def} then confirms that the expected EP
of the full system is lower-bounded by the expected EP of any one of its constituent units, $\oo$, a result
first derived in~\cite{wolpert2020minimal}.

As shorthand replace $t_f$ with $t$, and then expand 
\eq{
\left\langle  -\ln p^{X|X_\oo}_{\vx(t)}(t) \right \rangle &= -\sum_x p_x(t) \left(\ln p_x(t) - \ln p_{x_\oo}(t)\right)
}
Therefore,
\eq{
\dfrac{d}{dt} \left\langle - \ln p^{X|X_\oo}_{\vx(t)}(t) \right \rangle &= 
			-\sum_{x,x'} K^{x'}_x(t) p_{x'}(t) \ln p_x(t) \nonumber \\
		&\; + 	\sum_{x_\oo,x'_\oo} K^{x'_\oo}_{x_\oo}(\oo;t) p_{x'_\oo}(t) \ln p_{x_\oo}(t)
\label{eq:30aaa}
}

In addition, the sum in \cref{eq:27a} is just the total heat flow \textit{from} the systems in $-\oo$ \textit{into} 
their respective heat baths, during the interval $[0, t]$, if the system follows trajectory $\vx$. Therefore 
the derivative with respect to $t$ of the expectation of that sum is just the expected heat flow rate at $t$
from those systems into their baths,
\eq{
-\sum_{x,x'} K^{x'}_x(-\oo; t) p_{x'}(t)\ln \left[\dfrac{K^{x}_{x'}(-\oo;t) } {K^{x'}_{x}(-\oo;t) }\right]  
\label{eq:31a}
}

Note as well that $K(t) = K(\oo;t) + K(-\oo;t)$. So if we add \cref{eq:31a} to \cref{eq:30aaa},
and use the fact that rate matrices are normalized, we get
\eq{
\dfrac{d \left\langle \chi^\oo \right\rangle}{dt} &= -\sum_{x,x'} K^{x'}_x(\oo;t) p_{x'}(t) \ln p_{x|x_\oo}(t) \nonumber \\
		&\qquad + \sum_{x,x'} K^{x'}_x(-\oo; t)p_{x'}(t) \ln \left[\dfrac{K^{x'}_x(-\oo;t) p_{x'}(t)}{K^{x}_{x'}(-\oo;t) p_{x}(t)}\right]
\label{eq:31aa}
}

In~\cite{wolpert2020minimal},  the first sum in \cref{eq:31aa} is called the ``windowed derivative'',
$\dfrac{d^\oo}{dt} S^{X|X_\oo}(t)$. 
Since $\oo$ is a unit, it is the (negative) of the derivative of the mutual
information between $X_\oo$ and $X_{-\oo}$, under a counterfactual rate matrix in which $x_{-\oo}$ is held
fixed. As discussed in~\cite{wolpert2020minimal}, by the data-processing inequality, this term is non-negative.

The second sum in \cref{eq:31aa} was written as $\left\langle \dot{\sigma^\NN}_{K(\NN\setminus \oo; t)} \right\rangle$ 
in~\cite{wolpert2020minimal}. Since it is the expected rate of EP for a properly normalized, 
counterfactual rate matrix, it too is non-negative. Therefore the full expectation $\dfrac{d \left\langle \chi^\oo \right\rangle}{dt}$
is non-decreasing in time.

This decomposition of $d \left \langle \sigma^\NN - \sigma^\oo \right \rangle / dt$ was first derived in~\cite{wolpert2020minimal}.
However, that derivation did not start from a trajectory-level definition of local and global EPs, as done here.

$ $

\section*{APPENDIX D: Proof of vector-valued DF}
\label{app:DFT_vector}
Proceeding in the usual way~\cite{van2015ensemble,esposito.harbola.2007,seifert2012stochastic}, we first want to calculate
\eq{
\ln \left[\dfrac{\vP(\vx_{\AAA})} {\tilde{\vP}(\tilde{\vx}_{\AAA})} \right]
\label{eq:63}
}
where $\vx_\AAA$ is the trajectory of the states of the systems in $\cup_i \AAA_i$.
Note that any transition in $\vx_\AAA$ involves the change in the state of a single system,
due to the fact that we have an MPP. 
Exploiting this, we can parallel
the development in App.\,A of~\cite{esposito.harbola.2007}, to reduce the expression in \cref{eq:63}
to a sum of two terms. The first term is a sum, over all transitions in $\vx_\AAA$, of
the log of the ratio of two associated entries in the rate matrix of the system that changes state in that 
transition.\footnote{If there are no chemical reservoirs, then since each system is coupled to its own heat bath,  
we can uniquely identify which bath was involved in each state transition in any given $\vx$ directly from $\vx$ itself.
This is not the case for trajectory-level analyses of systems which are
coupled to multiple mechanisms, e.g.,~\cite{esposito.harbola.2007}; to 
identify what bath is involved in each transition in that setting we need to know more than just $\vx$.}
Since a union of units is a unit, we can use \cref{eq:local_EF_unit} to show that
that first sum equals $-Q^{\AAA}(\vx_{\AAA})$, the total EF generated by those systems during the trajectory. 
The second term is just $\Delta s^{\AAA}(\vx_{\AAA})$. So by the definition of local EP, 
we have a DFT over trajectories, 
\eq{
\ln \left[\dfrac{\vP(\vx_{\AAA})} {\tilde{\vP}(\tilde{\vx}_{\AAA})} \right] &= \sigma^{\AAA}(\vx_{\AAA})
\label{eq:31}
}

Similarly, for any single unit $\oo \in \{\AAA_i\}$,
\eq{
\ln \left[\dfrac{\vP(\vx_\oo)}{\tilde{\vP}(\tilde{\vx}_\oo)}\right] &= \sigma^\oo(\vx_\oo)
\label{eq:40b}
}
Therefore,
paralleling~\cite{van2015ensemble}, we can combine \cref{eq:31,eq:40b} to get a DFT for the probability density function of values of
$\vec{\sigma}^\AAA(\vx_{\AAA})$:
\eq{
\vP(\vec{\sigma}^\AAA) &= \int \mathcal{D}\vx_{\AAA} \, \vP(\vx_{\AAA}) 
		\prod_{\oo \in \AAA} \delta\left(\sigma^\oo - \ln \left[\dfrac{\vP(\vx_\oo)}{\tilde{ {\vP}}(\tilde{ \vx}_\oo)}\right]\right) \\
	& = e^{\sigma^{\AAA}} \int \mathcal{D}\vx_{\AAA} \, \tilde{ {\vP}}(\tilde{ \vx}_{\AAA}) 
		\prod_{\oo \in \AAA} \delta\left(\sigma^\oo - \ln \left[\dfrac{\vP(\vx_\oo)}{\tilde{ {\vP}}(\tilde{ \vx}_\oo)}\right]\right) \\
	& = e^{\sigma^{\AAA}} \int \mathcal{D}\tilde{\vx}_{\AAA}\, \tilde{ {\vP}}(\tilde{ \vx}_{\AAA}) 
		\prod_{\oo \in \AAA} \delta\left(-\sigma^\oo - \ln \left[\dfrac{\tilde{\vP}(\tilde{\vx}_\oo)}{{ {\vP}}({ \vx}_\oo)}\right]\right) \\
	& := e^{\sigma^{\AAA}} \tilde{\vP}(-\vec{\sigma}^\AAA)
}
Rewriting, we have shown that 
\eq{
\ln \left[\dfrac{\vP(\vec{\sigma}^\AAA)}{ \tilde{\vP}(-\vec{\sigma}^\AAA)}\right]  &= \sigma^{\AAA}
\label{eq:50_app}
}
which establishes the claim. 

As always, the reader should bear in mind that
${ \tilde{\vP}(-\vec{\sigma}^\AAA)}$ is the probability of a reverse trajectory $\tilde{ \vx}$, generated under $\tilde{ {\vP}}$,
such that if \textit{its} inverse, $\vx$, had been generated in the forward process, it would have resulted in 
$\vec{\sigma}^\AAA$. In turn, $\vec{\sigma}^\AAA$
is the vector of EPs of the trajectory as measured using the formula for \textit{forward}-process 
EPs~\cite{van2015ensemble}. So in particular, it involves EPs defined in terms of the drop in stochastic
entropy of the forward process, \textit{not} of the reverse process.

As an aside, note that ``vector-valued fluctuation theorems'' were derived previously in~\cite{garcia2010unifying,garcia2012joint}, using
similar reasoning. However, those FTs did not involve vectors of local EPs. Rather than held for any choice of vector $(A_1(\vx,K(t)), A_2(\vx,K(t)), \ldots)$
such that: i) global EP of any trajectory is $\sigma^\NN(\vx) = \sum_i A_i(\vx,K)$; ii) each component
$A_i(\vx, K(t))$ is odd if you time-reverse the trajectory $\vx$ and the sequence of rate matrices, $K(t)$.
The first property does not hold in general for the vector of local EPs.


$ $

\section*{APPENDIX E: Implications of the vector-valued DFT}
\label{app:other_implications}


In this appendix I derive \cref{eq:22} in the main text. I then present a simple example of the
claim made in the main text that for broad classes of MPPs, including the one depicted in \cref{fig:1},
$I\left(\vP(\sigma^{AB}, \sigma^{BC}) \; ||\; \tilde{\vP}(-\sigma^{AB}, -\sigma^{BC})\right) \ge 0$.
I follow this by briefly discussing some other implications of the
vector-valued DFT.


\subsection*{Decomposition of expected global EP involving multi-divergence}

As in the main text, let $\AAA$ be any set of units, not necessarily a full unit structure, but
including the unit $\cup_i \AAA_i$.
$\sigma^{ \AAA}$ is a single-valued function of $\vec{\sigma}^\AAA$, given simply by projecting $\vec{\sigma}^\AAA$
down to the associated component. Therefore
taking the average of both sides of the vector-valued DFT, \cref{eq:50}, over all $\vec{\sigma}^\AAA$ establishes that
\eq{
\left \langle \sigma^{ \AAA} \right \rangle &= D\left(\vP(\vec{\sigma}^\AAA) \,||\, \tilde{\vP}(-\vec{\sigma}^\AAA) \right)
\label{eq:40a}
}
(Note that this equality involves the KL divergence between two distributions over \textit{vectors} of EPs, and
should not be confused with the equality involving the KL divergence between two distributions over \textit{scalars} of EPs,
$\left \langle \sigma^\NN \right \rangle = D\left(\vP({\sigma^\NN}) \,||\, \tilde{\vP}(-{\sigma^\NN}) \right)$.)

Next, since any union of units is a unit, \cref{eq:23d} tells us that 
$\left \langle \sigma^\NN \right \rangle \ge \langle \sigma^{ \AAA} \rangle$. If we plug \cref{eq:40a} into this and then
add and subtract $\sum_i \left\langle \sigma^{\AAA_i} \right\rangle$ on the RHS, 
we derive \cref{eq:22} in the  main text:
\eq{
\left\langle \sigma^\NN \right\rangle &\ge \sum_{i} \left\langle \sigma^{\AAA_i}  \right\rangle 
		+ D\left(\vP(\vec{\sigma}^\AAA) \; ||\; \tilde{\vP}(-\vec{\sigma}^\AAA)\right)  	- 	\sum_{i} D\left(\vP({\sigma}^{\AAA_i}) \;||\; \tilde{\vP}(-\sigma^{\AAA}_i)\right) \\
&=  \sum_{i} \left\langle \sigma^{\AAA_i}  \right\rangle + 
		I\left(\vP(\vec{\sigma}^\AAA) \; ||\; \tilde{\vP}(-\vec{\sigma}^\AAA)\right)
\label{eq:e6}
}
Intuitively, the multi-divergence on the RHS of \cref{eq:e6} measures how much of the distance
between $\vP(\vec{\sigma}^\AAA)$ and  $\tilde{ {\vP}}(-\vec{\sigma}^\AAA)$ arises from the correlations
among the variables $\{\sigma^{\AAA_i}\}$, in addition to the contribution from the marginal distributions of each variable
considered separately.

%
%
%
%

\subsection*{Example of non-negativity of multi-divergence}

In this appendix I 
illustrate how to calculate multi-divergence, using the multi-divergence
$I\left(\vP(\sigma^{AB}, \sigma^{BC}) \; ||\; \tilde{\vP}(-\sigma^{AB}, -\sigma^{BC} )\right) $
from the ligand-sensing scenario. (So there are two units in $\AAA$, namely $AB$ and $BC$.)

Consider instances of this scenario obeying the following assumptions:
\begin{enumerate}
\item The rate matrices are time-homogeneous. 
\item The Hamiltonian is uniform and unchanging.
\item The distribution that the system starts with has the property that the map from $\left(p_{x_A, x_B}(0) \;,\; p_{x_C, x_B}(0) \right)$ to $(x_A, x_B, x_C)$
is single-valued. So  with $|X|$ the number of states in the joint system,  the vector-valued function $F$ defined by 
\eq{
F(x_A, x_B, x_C) := \left(\ln p_{x_A, x_B}(0) + \ln |X^{AB}|,\; \ln p_{x_C, x_B}(0) + \ln |X^{BC}| \right)
}
is invertible. Similarly, the functions $F_{AB}(x_A, x_B) := \ln p_{x_A, x_B}(0) + \ln |X^{AB}|$ and
$F_{BC}(x_C, x_B) := \ln p_{x_C, x_B}(0) + \ln |X^{BC}|$ are assumed invertible.

\item $t_f$ is large enough on the scale of the rate matrices so that during the MPP
the joint system relaxes from its (perhaps highly) non-equilibrium initial distribution to being arbitrarily
close to the uniform distribution, and then evolves no further. 
\end{enumerate}

When the second condition is met, i.e., the Hamiltonian is uniform for all $t$,
there is no EF into any of the reservoirs along any allowed trajectory
of states. So the EP generated by any unit $\oo$ along any allowed trajectory $\vx$
that goes from $x(0)$ to $x(t_f)$ is $- \Delta  \ln p_{x_\oo} = \ln p_{x_\oo}(0) - \ln p_{x_\oo}(t_f)$. In addition,
by the fourth condition, $p_x(t_f)$ is uniform.
This means, for example, that
\eq{
{\sigma}^{AB}(\vx) &= \ln p_{x_{AB}}(0) - \ln p_{x_{AB}}(t_f)  \nonumber \\
	&=  \ln p_{x_{AB}}(0) + \ln |X^{AB}|
}

Note that the reverse process is identical to the forward process, by the second condition.
Since the ending distribution is by hypothesis a fixed point
of the forward process, this means that the ending distribution is also a fixed point of the reverse process. So at the beginning of the
reverse process the distribution over $x$, $\tilde{p}_x(t_f)$ is uniform. Since the rate matrix is time-homogeneous,
the distribution at the end of the reverse process, $\tilde{p}_x(2 t_f)$, must also be uniform. 

Next recall the comment below \cref{eq:50_app} about how to interpret reverse process probabilities of negative
EP values. Using the assumed invertibility of $F$ and combining, this gives

\eq{
& {\tilde{\vP}}(-{\sigma}^{AB},- \sigma^{BC}) \nonumber \\
&\qquad = \sum_{x_{t_f},x_{2t_f}}  \tilde{p}(x_{2t_f} | x_{t_f}) \tilde{p}_x(t_f) \times
	\delta\left[{\sigma}^{AB},  \ln p_0(x^{AB}_{2t_f}) + \ln |X^{AB}|\right]  \times \delta\left[{\sigma}^{BC},  \ln p_0(x^{BC}_{2t_f}) + \ln |X^{BC}|\right] \\
	&\qquad= |X|^{-1}\sum_{x_{2t_f}} 
		\delta\left[{\sigma}^{AB},  \ln p_0(x^{AB}_{2t_f}) + \ln |X^{AB}|\right]  \times \delta\left[{\sigma}^{BC},  \ln p_0(x^{BC}_{2t_f}) + \ln |X^{BC}|\right]
\label{eq:e7}
}
where $|X|$ is the number of states of the joint system,
$\tilde{p}(x_{2t_f} | x_{t_f})$ is defined as the probability that if the reverse process is run starting from $x_{t_f}$
it ends at $x_{2t_f}$, and with some abuse of notation,
$p_t(x)$ is defined as the probability of state $x$ at time $t$ in the forward process. 

Again invoking the assumed invertibility
of $F$, we can evaluate \cref{eq:e7} to get
\eq{
{\tilde{\vP}}(-{\sigma}^{AB},- \sigma^{BC}) &= |X|^{-1} \sum_{x}  \delta[F(x), (\sigma^{AB}, \sigma^{BC})] 
}
So ${\tilde{\vP}}(-{\sigma}^{AB},- \sigma^{BC})$ has the same value, $|X|^{-1}$,
for all $|X|$ pairs $({\sigma}^{AB}, \sigma^{BC})$ that occur with nonzero probability under the forward process.

This allows us to evaluate
\eq{
D\left(\vP({\sigma}^{AB}, \sigma^{BC}) \; ||\; {\tilde{\vP}}(-{\sigma}^{AB},- \sigma^{BC})\right)
	&= \sum_{\sigma^{AB}, \sigma^{BC}} \vP(\sigma^{AB}, \sigma^{BC}) 
			  \left(	\ln{\vP(\sigma^{AB}, \sigma^{BC})} + \ln |X| \right) \nonumber \\
& = -S\left(\vP(\sigma^{AB}, \sigma^{BC})\right)  + \ln |X|
}
Similarly,
\eq{
D\left(\vP({\sigma}^{AB}) \; ||\; {\tilde{\vP}}(-{\sigma}^{AB})\right) &= 
\sum_{\sigma^{AB}} \vP(\sigma^{AB}) 
				\ln \left(\dfrac{\vP(\sigma^{AB})}{\tilde{\vP}(-\sigma^{AB})}\right) \nonumber \\
	& = -S\left(\vP(\sigma^{AB})\right) +  \ln |X^{AB}|
}
and
\eq{
D\left(\vP({\sigma}^{BC}) \; ||\; {\tilde{\vP}}(-{\sigma}^{BC})\right) &= 
\sum_{\sigma^{BC}} \vP(\sigma^{BC}) 
				\ln \left(\dfrac{\vP(\sigma^{BC})}{\tilde{\vP}(-\sigma^{BC})}\right) \nonumber \\
	& = -S\left(\vP(\sigma^{BC})\right) +  \ln |X^{BC}|
}
Combining, we see that when the four conditions given above are met, the multi-divergence 
is just the mutual information between $\sigma^{AB}$ and $\sigma^{BC}$, plus the positive quantity
$\ln |X^{BC}| - \ln |X^{B}|$.
%
%
So the multi-divergence
$I\left(\vP(\sigma^{AB}, \sigma^{BC}) \; ||\; \tilde{\vP}(-\sigma^{AB}, -\sigma^{BC} )\right)$ 
is non-negative in this situation. 

The same kind of reasoning can be
extended to establish that many MPPs other than the one considered here also must
have non-negative multi-divergence. (In larger MPPs,
rather than invoke non-negativity of mutual information, as done here, one invokes non-negativity of total correlation.)

\begin{example}
To illustrate how the four conditions result in non-negative multi-divergence,
in the rest of this subsection I calculate the multi-divergence explicitly, for an explicitly specified
initial distribution. 

To begin, use the chain rule for relative entropy to expand 
the multi-divergence in the discussion of the ligand-sensing example  in the main text, as
\eq{
&I\left(\vP(\sigma^{AB}, \sigma^{BC}) \; ||\; \tilde{\vP}(-\sigma^{AB}, -\sigma^{BC} )\right) =   \nonumber \\
& \qquad D\left(\vP({\sigma}^{AB} \,|\, \sigma^{BC}) \; ||\; { \tilde{\vP}}(-{\sigma}^{AB} \,|\, -\sigma^{BC})\right) - 
		D\left(\vP({\sigma}^{AB}) \; ||\;  \tilde{\vP}(-{\sigma}^{AB})\right) 
\label{eq:e8a}
}

Next, suppose that systems A and C have the same number of states, i.e., $|X_A| = |X_C|$. Assume further that those
two systems each have a single reservoir, with the same temperature, and that the two rate matrices $K(C)$ and
$K(A)$ are identical (under the interchange $x_A \leftrightarrow x_C$). Suppose as well that $B$ is in a stationary state.
In addition choose
\eq{
p_{x_A, x_B, x_C}(0) =\dfrac{ f(x_A) \delta(x_A, x_C)}{|X^B|}
\label{eq:e13}
}
for some invertible normalized distribution $f(.)$. So $p_{x_B | x_A, x_C}(0)$ is uniform, independent of the
values of $x_A$ and $x_C$. 

As above, we assume that the Hamiltonian is uniform and unchanging, the rate matrices are time-homogeneous,
and the system ends up at $t_f$ in a uniform distribution that is a fixed point of the dynamics.
So again, no work is done on the system, and therefore the EP along a trajectory is the associated drop in the stochastic entropy. 
%
Moreover, since the ending distribution is uniform, \cref{eq:e13} gives
\eq{
\sigma^{AB}(\vx) &= -\Delta \ln p_{x_{AB}}(\vx) \nonumber \\
	&=  \ln |X^{AB}| + \ln f(\vx_A(0)) - \ln |X^B|   \nonumber \\
	&=  \ln |X^{A}| + \ln f(\vx_A(0))
\label{eq:e9}
}
and similarly for $\sigma^{BC}(\vx)$. 

Next, note that since $\vx_A(0) = \vx_C(0)$ with probability $1$, \cref{eq:e9} means that
$\sigma^{AB} = \sigma^{BC}$ with probability $1$. Therefore, since $f$ is invertible, using \cref{eq:e13} gives
\eq{
\vP({\sigma}^{AB}, \sigma^{BC}) &= \delta ({\sigma}^{AB}, \sigma^{BC}) 
	\sum_{x_A} f(x_A)  \delta[{\sigma}^{AB},  \ln |X^{A}| + \ln f(x_A)]     \nonumber \\
	 &= \delta ({\sigma}^{AB}, \sigma^{BC})e^{ \sigma^{AB} } \;/ \; |X^A|
}
for all pairs $({\sigma}^{AB}, \sigma^{BC})$ that occur with nonzero probability under the forward process.
This in turn means that
\eq{
\vP({\sigma}^{AB}) &= e^{  \sigma^{AB} }  \;/ \; |X^A|
\label{eq:e12}
}
for all values $\sigma^{AB}$ that occur with nonzero probability under the forward process.

On the other hand, given that the reverse process ends in a uniform distribution, by \cref{eq:e9} and its analog for $\sigma^{BC}$,
\eq{
&\tilde{\vP}(-{\sigma}^{AB}, -\sigma^{BC}) = |X^A|^{-2}  \nonumber \\
&\qquad \times \sum_{x_A,x_C} \delta\left(\sigma^{AB}, \ln |X^A| + \ln f(x_A)\right)
					 \delta\left(\sigma^{BC}, \ln |X^A| + \ln f(x_C)\right)
}
(where use was made of the fact that $|X^A| = |X^C|$). Due to the invertibility of $f$,
this distribution has the same value, $|X^A|^{-1}$,
for all $|X|$ pairs $({\sigma}^{AB}, \sigma^{BC})$ that occur with nonzero probability under the forward process. So
%
\eq{
\tilde{\vP}(-{\sigma}^{AB}) &= \sum_{x_A} \delta\left(\sigma^{AB}, \ln |X^A| + \ln f(x_A)\right) \;/\;  |X^A|
}

Plugging these formulas into \cref{eq:e8a} gives
\eq{
&D\left(\vP({\sigma}^{AB}, \sigma^{BC}) \; ||\; {\tilde{\vP}}(-{\sigma}^{AB},- \sigma^{BC})\right)
			   \nonumber \\
&\qquad\qquad\qquad=  -\sum_{\sigma^{AB}}  \dfrac{e^{  \sigma^{AB} }}{|X^A| }\ln \left(\dfrac{|X^A|^{-1}}{ e^{ \sigma^{AB} }}\right)
							\nonumber \\
&\qquad\qquad\qquad=   \sum_{\sigma^{AB}} \dfrac{e^{  \sigma^{AB} }}{|X^A| }\left[\sigma^{AB} +  \ln |X^A| \right]
}
and
\eq{
&D\left(\vP({\sigma}^{AB}) \; ||\; {\tilde{\vP}}(-{\sigma}^{AB})\right) 	 =   
			 \sum_{\sigma^{AB}} \dfrac{e^{  \sigma^{AB} }}{|X^A| }\sigma^{AB} 
}
Combining and using \cref{eq:e12} gives
\eq{
&I\left(\vP({\sigma}^{AB}, \sigma^{BC}) \; ||\; {\tilde{\vP}}(-{\sigma}^{AB}, -\sigma^{BC})\right) 
		= \ln|X^A| \;-\; \langle \sigma^{AB} \rangle 
}
This is always non-negative since the maximal value of $\sigma^{AB}$ under any $f(.)$ is $\ln |X^A|$.
This establishes the claim.

\end{example}

\subsection*{Other implications of vector-valued DFT}


\cref{eq:40a} can be elaborated in several way. 
Let $\BBB$ be any subset of the units in $\AAA$. Using the chain-rule for KL divergence to expand
the RHS of \cref{eq:40a}, and then applying \cref{eq:40a} again, this time
with $\AAA$ replaced by $\BBB$, gives
\eq{
\label{eq:41}
\left \langle \sigma^{ \AAA} \right \rangle & =  \left \langle \sigma^{\BBB} \right \rangle +  
	D\left(\vP(\vec{\sigma}^\AAA \,|\, \vec{\sigma}^\BBB) \,||\,  \tilde{\vP}(\vec{\sigma}^\AAA  \,|\, \vec{\sigma}^\BBB) \right) 
}
So the difference in expected total EPs of $\AAA$ and $\BBB$ exactly equals the conditional KL-divergence between the associated EP vectors.

We can also apply the kind of reasoning that led from the vector-valued DFT to \cref{eq:40a}, but after
plugging \cref{eq:global_EP_decomp_in_ex}
in the main text  into 
the conditional DFT.
This shows that for any value of $\sigma^\oo$ with nonzero probability,
\eq{
& \widehat{\sum_{\oo'\in {\NN^*}}}  \left\langle  \sigma^{\oo'} \,\bigg|\, \sigma^\oo \right\rangle  - \left\langle \Delta \II^{{\NN^*}}  \,\bigg|\, \sigma^\oo \right\rangle  \nonumber \\
	&\qquad=   \sigma^\oo + D\left(\vP(\vec{\sigma}^{\NN^*} \,\downharpoonright\, \sigma^\oo) 
						\,||\, \tilde{\vP}(-\vec{\sigma}^{\NN^*} \,\downharpoonright\, -\sigma^\oo) \right)
}
(where $\downharpoonright$ indicates conditioning on a specific value of a random
variable rather than averaging over those values, as in the conventional definition of conditional relative entropy).
Similarly, the conditional DFT associated with \cref{eq:32c}
shows that for any value of $\sigma^\oo$ with nonzero probability,
\eq{
\left\langle \chi^\oo| \sigma^\oo \right\rangle &= D\left(\vP(\vec{\sigma}^{\NN^*} \,\downharpoonright\, \sigma^\oo) 
						\,||\, \tilde{\vP}(-\vec{\sigma}^{\NN^*} \,\downharpoonright\, -\sigma^\oo) \right)
}


$ $

\section*{APPENDIX F: Sufficient conditions for the conditional mutual information not to increase}
\label{app:conditional_independence_rule}

If $C$ observes $B$ continually, then it may gain some information about the trajectory of $B$
that is not necessarily captured by $B$'s final state.
That information however would tell $C$ something about the ending state
of $A$, if $A$ had also observed the entire trajectory of $B$. 
This phenomenon works to increase $I(A ; C \,|\, B)$ during the process.
On the other hand, in the extreme case where neither $A$ nor $C$ observes
$B$, and the three systems are initially independent of one
another, by the data-processing inequality the mutual information between
$A$ and $C$ will drop during  the process, and therefore so will 
 $\langle \Delta I(A ; C \,|, B)$. So in general, $\langle \Delta I(A ; C \,|, B)$ can be either negative or positive. 
In this appendix I present some simple sufficient conditions for it to be non-positive.

First, note that if both $p_{x_A,x_C \,|\, x_B}(0) \propto \delta(x_A, x_C)$ and
$p_{x_A,x_C}(t_f) = \delta(x_A, x')\delta(x_C,x')$ for some special $x'$,
then $ \Delta I(A; C \,|\, B)  \le 0$. (See~\cref{app:conditional_independence_rule} 
for other cases.) There are also many cases where $ \Delta I(A; C \,|\, B)  > 0$. For example, 
since conditional mutual information is non-negative, that is generically the case if 
$ I(A; C \,|\, B)(0)  =0$ and $x_B$ evolves non-deterministically.

In addition, in App.\,C in~\cite{kolchinsky2020entropy}, it was shown that if any initial distribution in
which $A$ and $C$ are conditionally independent given $B$
is mapped by the MPP to  a final distribution with the same property, then the MPP cannot increase
$I(A ; C \;|\; B)$, no matter the initial distribution actually is. Loosely speaking,
if the MPP ``conserves conditional independence of $A$ and $C$ given $B$'',
then it cannot increase $I(A ; C \;|\; B)$.

The claim was originally proven as part of a complicated
analysis. To derive it more directly, first note that for any joint distribution $p(x_A, x_B, x_C)$,
\eq{
\langle \Delta I_p(A; C \,|\, B) \rangle &= D(p_{X_{ABC}} \;||\; p_{X_{A|B}} p_{X_{C|B}} p_{X_B})
}
where $D(\cdot \;||\; .)$ is relative entropy. rite 
the conditional distribution of the entire MPP as $\Phi(\cdot)$, so 
for any initial distribution $q$, the associated ending distribution is $q(t_f) = \Phi \left(q(0)\right)$.
As shorthand, write $p' = \Phi(p_{X_{ABC}}(0))$.

Since by hypothesis the MPP sends distributions where $x_A$ and $x_C$ are conditionally independent
to distributions with the same property, there must be some distribution $q$ such that
\eq{
\Phi \left(p_{x_{A|B}}(0)p_{x_{C|B}}(0) p_{x_{B}}(0)\right) &= q_{x_{A|B}}(t_f)q_{x_{C|B}}(t_f)q_{x_{B}}(t_f)
}
In addition, since the overall MPP is a discrete time Markov chain, the chain rule for relative
entropy applies. As a result we can expand
\eq{
0 &\leq D\left(p_{A B C} \| p_{A \mid B} p_{C \mid B} p_{B}\right)-D\left(p_{A B C}^{\prime} \| \Phi\left(p_{A \mid B} p_{C \mid B} p_{B}\right)\right) \\
&=D\left(p_{A B C} \| p_{A \mid B} p_{C \mid B} p_{B}\right)-D\left(p_{A B C}^{\prime} \| q_{A \mid B} q_{C \mid B} q_{B}\right) \\
&=D\left(p_{A B C} \| p_{A \mid B} p_{C \mid B} p_{B}\right) - D\left(p_{A B C}^{\prime} \| p_{A \mid B}^{\prime} p_{C \mid B}^{\prime} p_{B}^{\prime}\right) \nonumber \\
&\qquad\qquad -D\left(p_{A \mid B}^{\prime} p_{C \mid B}^{\prime} p_{B}^{\prime} \| q_{A \mid B} q_{C \mid B} q_{B}\right) \\
&\leq D\left(p_{A B C} \| p_{A \mid B} p_{C \mid B} p_{B}\right)-D\left(p_{A B C}^{\prime} \| p_{A \mid B}^{\prime} p_{C \mid B}^{\prime} p_{B}^{\prime}\right)
}
This establishes the claim.

As an illustration, the MPP 
conserves conditional independence of $A$ and $C$ given $B$ (and so by the claim just
established, $\Delta I(A ; C \;|\; B) \le 0$) so long as $x_B$ evolves in an invertible deterministic manner during the MPP.
So in particular, i$\Delta I(A ; C \;|\; B) \le 0$ if $x_B$ does not change its state during the MPP.

%
To see this, as shorthand
write $x_A(t_f) = x'_A, x_A(0) = x_A$, and similarly for systems $B$ and $C$. Write the entire joint
trajectory as $\vx$, as usual. Under the hypothesis that $x_B$ evolves
deterministically, we can write $\vx_B(t) = V_{x_B}(t)$ for some
function $V$. Since that deterministic dynamics of $B$ is invertible, we can also write $V_{x_B}(t) = V'_{x'_B}(t)$
for some function $V'$, and can write $x_B = f(x'_B)$
for some invertible function $f$.

Using this notation, if the initial
distribution $A$ is conditionally independent of $C$ given $B$, then due to the unit structure we can expand
\eq{
p(x'_A, x'_B, x'_C) &= \sum_{x_A, x_B, x_C} \int \mathcal{D}\vx_A \mathcal{D}\vx_B \mathcal{D}\vx_C \; \delta(\vx_A(t_f), x'_A)
		 \delta(\vx_B(t_f), x'_B)  \delta(\vx_C(t_f), x'_C)  									 \nonumber \\
	&\qquad \qquad\qquad \qquad\qquad p(\vx_A \;|\; \vx_B, x_A) p(\vx_C \;|\; \vx_B, x_C) p(\vx_B  \;|\; x_B)
			p(x_A \;|\; x_B) p(x_C \;|\; x_B) p(x_B)										 \nonumber \\
	&= \sum_{x_A, x_B, x_C} \int \mathcal{D}\vx_A  \mathcal{D}\vx_C \; \delta(\vx_A(t_f), x'_A)
		 \delta(\vx_B(t_f), x'_B)  \delta(\vx_C(t_f), x'_C)   						\nonumber \\
	&\qquad \qquad\qquad \qquad\qquad p(\vx_A \;|\; V_{x_B}, x_A) p(\vx_C \;|\; V_{x_B}, x_C) 
			p(x_A \;|\; x_B) p(x_C \;|\; x_B) p(x_B)										 \nonumber \\
	&= \sum_{x_A, x_C}
p(x'_A \;|\; V'_{x'_B}, x_A) p(x'_C \;|\; V'_{x'_B}, x_C) 
			p(x_A \;|\; f(x_B')) p(x_C \;|\; f(x'_B)) p(f(x'_B))										 \nonumber \\
	&= 
p\left(x'_A \;|\; V'_{x'_B}, f(x'_B) \right) p\left(x'_C \;|\; V'_{x'_B}, f(x'_B) \right) p\left(f(x'_B)\right)	
}
which establishes that under the ending distribution, $A$ and $C$ are conditionally independent given $B$, as claimed.


$ $

\section*{APPENDIX G: Cases where subsystem LDB is exact under global LDB}
\label{app:exact_ldb}

In this paper I assume that \cref{eq:nldb} holds, and that 
all fluctuations in the state of system $i$ are due to exchanges with its reservoir(s), and that the amounts
of energy in such exchanges are given by the associated changes in the value of $i$'s  local 
Hamiltonian. (This is the precise definition of SLDB). In many scenarios SLDB will only hold
to very high accuracy. In particular, often in the literature the statement is made that a first system
reacts to the state of a second one, but that there is no ``back-action'' of the second system also 
reacting to the state of the first one~\cite{hartich_sensory_2016}. The validity
of this approximation can be semi-formally justified by supposing that there is a much larger range of 
energy values of the second system than the first~\cite{verley_work_2014}. 

In this appendix I present some sufficient conditions for SLDB to hold exactly. Then in \cref{app:approx_ldb}, I present  a
more rigorous version of the kind of reasoning invoked in~\cite{verley_work_2014}, establishing conditions
for SLDB to hold to high accuracy. 

\subsection*{Hamiltonian stubs}

Rather than start with a set of rate matrices, one per system, start with an additive decomposition of
the global Hamiltonian into a set of $D$ different terms, where each term $d \in \{1, \ldots, D\}$ only depends on the joint
state of the systems in some associated subset of $\NN$. Specifically, for each $d \in \{1, \ldots, D\}$,
choose some arbitrary set ${m(d)} \subseteq \NN$, along with an associated arbitrary \textbf{Hamiltonian stub}, $h_{d, t}(x_{m(d)})$,
and define the global Hamiltonian to be a sum of the stubs:
\eq{
H_x(t) = \sum_{d=1}^D h_{d, t}(x_{m(d)})
\label{eq:global_hamiltonian}
}
Next, for all systems $i$, define the associated neighborhood and local Hamiltonian by
\eq{
r(i) &:= \cup_{d : i \in m(d)} m(d) \\
H_{x}(i; t) &:= H_{x_{r(i)}}(i; t) \\
	&:= \sum_{d : i \in m(d)} h_{d, t}(x_{m(d)})
\label{eq:local_hamiltonian}
}

For an MPP with this global Hamiltonian, global, exact LDB says that fluctuations in the state of system $i$ due
to its reservoirs are governed by a Boltzmann distribution with Hamiltonian $H_{x}(i; t)$. So SLDB holds,
as claimed.

Note that the global Hamiltonian does not equal the sum of the local Hamiltonians in general, i.e., 
it may be that
\eq{
\sum_{d=1}^D h_{d, t}(x_{m(d)}) \ne \sum_i \sum_{d : i \in m(d)} h_{d, t}(x_{m(d)})
}
On the other hand, since in an MPP only one system changes state at a time,
the total heat flow to the reservoirs of all the systems along any particular trajectory $\vx$
\textit{is} the sum of the associated changes in the values of the local Hamiltonians (which are given in \cref{eq:local_hamiltonian}). 
For the same reason, the total change in the global Hamiltonian along any particular trajectory $\vx$
that involve the changes of the state of system $i$ --- the total heat flow into the system from the reservoirs
of system $i$ --- are given by the sum of the changes of the local Hamiltonian $H_{x}(i;t)$ along that trajectory.

\subsection*{Example involving diffusion between organelles}

 Consider a pair of systems, $A, C$, with an intervening ``wire'', $b$,
that diffusively transports signals from $A$ to $C$, but not vice-versa. It is meant to be an abstraction
of what happens when one cell sends a signal to another, or one organelle within a cell sends a 
signal to another organelle, etc. So $C$ is observing $A$, but
$A$'s dynamics is independent of the state of $C$. 

To formalize this scenario, let $Y$ be some finite space with at least three elements, with a special element labeled $0$.
Let $X_A$ be a vector space, with components $\{x_{A}(i) : i = 1, \ldots\}$, where
$X_{A}(1) = Y$. We will interpret $x_A(1)$ as the ``emitting signal''. 
Let $X_b = Y^m$ for some $m \ge 3$, and write those $m$ components of $x_b$
as $x_b(1), x_b(2), \ldots x_b(m)$. Intuitively, the successive components of $b$ are successive positions on a physical wire, stretching
from $A$ to $C$; if $x_{b}(j) = 0$, then there is no signal at location $j$ on the wire. 
Finally, let $X_C$ also be relatively high-dimensional, with components $\{x_{C}(i) : i = 1, \ldots\}$,
where $X_{C}(1) = Y$. We will interpret $x_C(1)$ as the ``received signal''. 

Have four Hamiltonian stubs
which are combined as in \cref{eq:global_hamiltonian} to define that global Hamiltonian.
Also make the definitions in \cref{eq:local_hamiltonian}. So global LDB automatically
ensures SLDB. 
From now on, to reduce notation clutter, the time index $t$ will be implicit.

Choose $m(1) = \{A\}, m(2) = \{A, b, C\}$, $m(3) = \{b, C\}$ and $m(4) = \{C\}$. Therefore 
$r(A) = r(b) =  r(C) = \{A, b, C\}$.
Also  choose $h_3$ to only depend on the first component of $x_C$ and on the $m$'th component of $x_b$; 
choose $h_2$ to only depend on $x_b$ and the first components of $x_A$ and $x_C$; and choose
$h_1$ to only depend on $x_A$. So
we can write the third stub as $h_3(x_b(m), x_C)$j, the second stub as
$h_2(x_A(1), x_b, x_{C}(1))$, and the first stub as $h_1(x_A)$

Combining, the local Hamiltonians of the three subsystems are
\eq{
\label{eq:full_A_Hamiltonian}
H_{x_A, x_b, x_C(1)}(A) &= h_1(x_A) + h_2(x_A(1), x_b, x_{C}(1)) \\
H_{x_A(1), x_b, x_C(1)}(b) &=  h_2(x_A(1), x_b, x_{C}(1)) + h_3(x_b(m), x_C(1))   \\
		&:= \hat{h}_2(x_A(1), x_b, x_{C}(1))) 
 \\
H_{x_{A}(1), x_b(m), x_C(1)}(C) &= h_2(x_A(1), x_b, x_{C}(1)) \nonumber \\
	&\qquad \qquad + h_3(x_b(m), x_C(1))  + h_4(x_C) \\
	&= \hat{h}_2(x_A(1), x_b, x_{C}(1))) + h_4(x_C) 
\label{eq:full_C_Hamiltonian}
}

For simplicity assume that at most one signal can be on the wire at a given time. So  $h_2(x_A(1), x_b, x_{C}(1)) = \infty$
if more than one component of $x_b$ differs from $0$, and therefore the same is
true for  $\hat{h}_2(x_A(1), x_b, x_{C}(1))$ (assuming that $h_3$ is never negative-infinite). 
Assume that ${h}_2(x_A(1), x_b, x_{C}(1))$ has the same fixed value
for all other values of its arguments. This means that ${h}_2(x_A(1), x_b, x_{C}(1))$ is independent
of the values of $x_A(1)$ and $x_C(1)$.
So as far as the rate matrices of $A$ and $C$ are concerned, by LDB
we can rewrite \cref{eq:full_A_Hamiltonian,eq:full_C_Hamiltonian} as
\eq{
\label{eq:e36}
H_{x_A}(A) &= h_1(x_A) \\
H_{x_B, x_C}(C) &= h_3(x_b(m), x_C(1)) + h_4(x_C)
\label{eq:e37}
}
\cref{eq:e36} establishes that the dynamics of $A$ is autonomous, independent of the 
states of $B$ and $C$, as claimed. 

As a final requirement of the Hamiltonians, for all $x, x'$ such that $x_C(1) = x_b(m), x'_C(1) \ne x_b(m)$, take
\eq{
h_3\left(x_b(m), x_C(1)\right)  < h_3(x_b(m), x'_C(1))
\label{eq:e35a}
}
This imposes a bias on the rate matrix of $C$, for $x_C(1)$ to change
to equal $x_b(m)$. This bias is the source of the asymmetry between $A$ (which as elaborated below can emit signals
into the wire but never reacts to what's on the wire) and $C$ (which as elaborated below can absorb signals
from the wire). Note though that this stipulation concerning $h_3$ does not
change the fact that the local Hamiltonian of $b$ has the same fixed value so long as 
no more than one component of $x_b$ is nonzero. Phrased differently, we could change
$h_3$ arbitrarily, even getting rid of it, and so long as $h_2$ is changed in a corresponding
manner, $\hat{h}_2$ would not change, and therefore neither would the local Hamiltonian of $b$.

We impose extra restrictions on $K(b)$ in addition to LDB:
\begin{enumerate}
\item 
\label{item:constraint1}
Suppose both $x'_b$ and $x_b$ have exactly one nonzero component, with indices $j', j$,
respectively, and write
$x'_b(j') = y'$ and $x_b(j) = y$, where neither $y$ nor $y'$ equals $0$. Then 
\eq{
K^{x_A(1), x'_b, x_{C}(1)}_{x_A(1), x_b, x_{C}(1)}(b) = 0
} 
if $y' \ne y$. So the signal in the wire cannot spontaneously change.

If $y = y'$ however, and $|j-j'| = 1$,
then 
\eq{
K^{x_A(1), x'_b, x_{C}(1)}_{x_A(1), x_b, x_{C}(1)}(b; t) \ne 0
}
So the signal can diffuse (without changing) across the wire.

Note that in this case where $y = y'$ and $|j-j'| = 1$, LDB means that
$K^{x_A(1), x'_b, x_{C}(1)}_{x_A(1), x_b, x_{C}(1)}(b; t)$ must be the same whether $j = j'+1$ or $j = j'-1$, i.e., 
the signal can go either backwards or forwards between $A$ and $C$. 

We also require that if $x'_b = \vec{0}$, and $x_b(j) \ne 0$ for some $1 < j < m$, then
\eq{
K^{x_A(1), x'_b, x_{C}(1)}_{x_A(1), x_b, x_{C}(1)}(b) = 0
} 
So a signal cannot spontaneously appear at a position between its two ends. (Which in turn means by LDB
that an already existing signal cannot spontaneously disappear from such a position in the wire.)
\item 
\label{item:constraint2}
If $x'_b = \vec{0}$ (i.e., all components of $x'_b$ equal $0$), and $x_b(1) = x_A(1) \ne 0$ then
\eq{
K^{x_A(1), x'_b, x_{C}(1)}_{x_A(1), x_b, x_{C}(1)}(b) \ne 0
\label{eq:e35}
}
So the wire can copy the signal in $x_A(1)$ into its first position.

Note that by LDB, the signal can even be ``re-absorbed'' back into $A$,
in the sense that while it is (still) in $x_A(1)$, it is no longer in $x_b(1)$, i.e., 
so that $x_b(1)$ fluctuates back to ${0}$ from some different value. (We might be able to remove this possibility
by appropriate modification of $h_2$.) 

However, if instead $x'_b = \vec{0}$ and $x_A(1) \ne 0$ but $x_b(1) \ne x_A(1)$, then 
\eq{
K^{x_A(1), x'_b, x_{C}(1)}_{x_A(1), x_b, x_{C}(1)}(b)= 0
}
So the wire cannot make a mistake when copying the signal in $x_A(1)$ into its first position.
\item 
\label{item:constraint3}
If $x'_b(m) = x_C(1)$ and $x_b = \vec{0}$, then
\eq{
K^{x_A(1), x'_b, x_{C}(1)}_{x_A(1), x_b, x_{C}(1)}(b) \ne 0
}
So if the signal in $x_b(m)$ has already been copied into $x_C(1)$ by $K(C)$ --- or serendipitously already exists
in $x_C(1)$ --- then $x_b(m)$ can be reset to $0$. Such a resetting will allow a new signal to enter
the wire from $A$.

Note that by LDB, if $x'_b = \vec{0}$, and $x_b(j) = 0$ for all $j < m$ while $x_b(m) = x_C(1)$, then 
\eq{
K^{x_A(1), x'_b, x_{C}(1)}_{x_A(1), x_b, x_{C}(1)}(b) \ne 0
} 
So it is possible for a nonzero signal to diffuse back into the wire from $x_C(1)$.
However, due to \cref{eq:e35a}, 
once $x_b(m)$ is reset to $0$, $x_C(1)$ is likely to follow suite, and also change to $0$. That would remove the signal 
from $C$, and so stop it from diffusing back into the wire. (N.b., $A$ does not have this kind of coupling with the wire,
only $C$.)
\end{enumerate}
Our final rule is that all other terms in the matrix $K(b)$ that are not set by either LDB or normalization equal $0$. So in particular,
if $x'_b(m) \ne x_C(1)$ and $x'_b(m) \ne 0$ while $x_b = \vec{0}$, then
\eq{
K^{x_A(1), x'_b, x_{C}(1)}_{x_A(1), x_b, x_{C}(1)}(b) = 0
}
This means that the wire cannot be reset until the signal in it has been copied into $x_C(1)$.

Several things about this scenario are worth noting. First,
the set of rate matrix constraints on $K(b)$ is consistent
with LDB, since all states of the wire with nonzero probability have the same energy value, $E$.
In particular, LDB can be consistent with \cref{eq:e35}; it simply means that if the wire has a single nonzero entry,
in its first location, which happens to equal $x_A(1)$, then the wire can ``fluctuate'' into the state $\vec{0}$,
losing all information.
Similarly, $H_{x}(b)$ is independent of the value of $x_A$, and so its value doesn't change depending on how
$x_A$ is related to $x_B$. Nonetheless, the associated rate matrix $K^{x'}_x(b)$ does
change depending on the relation between $x'_A(1)$ and $x'_b(1)$.

Second, if $x_b = \vec{0}$, two things can happen: either $x_{C}(1)$ gets copied into $x_{b}(m)$ (due to
LDB and constraint (c) on $K(b)$) or $x_{A}(1)$ gets copied into $x_b(1)$ (by constraint (b) on $K(b)$).
The first of those possibilities means that even though there is no back-action of $b$ on $A$, there \textit{is} back-action of $C$ on $b$, i.e.,
information can propagate from $C$ into $b$. Finally, note that there are two units in this
scenario: $\{A\}$ and $\{A, b, C\}$. 

$ $

\section*{APPENDIX H: Subsystem LDB as an approximation to global LDB}
\label{app:approx_ldb}

$ $

For some unit structures, requiring SLDB to hold to high accuracy 
imposes restrictions on the possible relationship among the 
relative scales of the underling Hamiltonians, chemical potentials and temperatures of the subsystems. 
This appendix illustrates some of these restrictions.

\subsection*{Global LDB}

To begin, note that the thermodynamic analysis in the main text, involving SLDB, only considers local, subsystem-specific Hamiltonians.
However, strictly speaking, local detailed balance is a restriction on the relationship between the rate matrices
and a \textbf{global Hamiltonian}, $H_x(t)$, which encompasses all the subsystems. More precisely,
since we have a multipartite system, we can decompose the \textbf{actual} full rate matrix as $\K^{x'}_x(t) = \sum_i \K^{x'}_x(i; t)$,
where for all $i$, $\K^{x'}_x(i; t) = 0$ if $x'_{-i} \ne x_{-i}$. 
Strict global LDB then tells us that for all $i, k, x, x'$, where $x'_{-i} = x_{-i}$, 
\eq{
&\ln \dfrac{\K^{x'}_{x}(i; k, t)}{\K^{x}_{x'}(i; k, t)} = \beta_i^k\left(\left[H_{x'}(t) - H_{x}(t)\right] + \mu_{i}^k\left[n_i^k(x_i) - n_i^k(x'_i)\right]\right)  
\label{eq:strict_LDB}
}

As discussed in the text, many stochastic thermodynamics models of one subsystem $A$ observing another subsystem $B$
assume that the dynamics of $B$
is independent of the state of $A$. Such an assumption is sometimes phrased as there being no
``back-action'' on the state of $B$ during the
observation~\cite{verley_work_2014,sagawa2012fluctuation,parrondo2015thermodynamics,horowitz2011designing,shiraishi_ito_sagawa_thermo_of_time_separation.2015,wachtler2016stochastic}. It is equivalent to assuming a simple unit structure governing the two subsystems, in which $B$ is one unit, and the pair $(A, B)$ is another.
In this appendix I investigate the relationship between this assumptions of no back-action and \cref{eq:strict_LDB},
and the implications of that relationship for the legitimacy of requiring SLDB.

First, it is important to note that there are instances of one subsystem observing another 
which do not exactly obey \cref{eq:strict_LDB} for \textit{any} (global) Hamiltonian $H$.
This is illustrated in the following example:

\begin{example}
Consider again the bipartite systems investigated in~\cite{barato_efficiency_2014,hartich_sensory_2016} that were described above,
which involve an internal subsystem $B$ and an independently evolving external subsystem $A$ that is observed by $B$. 
For simplicity assume there are no chemical potentials. 

In these systems, since $B$ observes $A$, the dynamics across $X_B$ at any given time must vary depending on the state of $X_A$
at that time $t$. (That's a minimal condition to be able to say that $B$ ``observes'' $A$.) That means
that the terms in the rate matrix of $B$ must vary depending on the state $x_A$. 
Suppose that in fact $B$ acts as a memory of the state of $A$: if we change the state of $A$,
we change the relative ratio of the associated equilibrium probabilities of the states of $B$. 
Strictly speaking,  LDB would then mean that there is a set of states $x_A, x'_A \ne x_A, x_B, x'_B \ne x_B$ such that
\eq{
H_{x_A, x_B'} - H_{x_A, x_B} \ne H_{x'_A, x_B'} - H_{x'_A, x_B}
}
for the {global} Hamiltonian $H_{x}$. (For clarity, the time index has been suppressed.) This would in turn mean that 
\eq{
H_{x_A, x'_B} - H_{x'_A, x'_B} \ne H_{x_A, x_B} - H_{x'_A, x_B}
}

\cref{eq:strict_LDB} would then require the rate matrix of subsystem $A$ to depend on the state $x_B$, assuming that $A$'s rate matrix is nonzero at $t$.
That is contrary to the assumption that $A$'s dynamics is independent of the state of $B$.
So the unit structure in~\cite{barato_efficiency_2014,hartich_sensory_2016} does not strictly obey LDB for \textit{any} global Hamiltonian.
In other words, it is not thermodynamically consistent.
\label{ex:1}
\end{example}


Even in cases like this example, it is legitimate to perform the thermodynamic
analysis using a rate matrix $K(t)$ that obeys SLDB for the given unit structure $\NN^*$ rather than using the actual rate matrix $\K(t)$ (which 
obeys global LDB but typically violates SLDB) if
that rate matrix $K(t)$ is extremely close to $\K(t)$. To make this more precise, write $\ovP(\vx)$ for the trajectory $\vx$ under actual
rate matrix $\K(t)$, for some initial distribution $p_x(0)$, and as before, write $\vP(\vx)$ for the trajectory $\vx$ under SLDB
rate matrix $K(t)$, for the same initial distribution $p_x(0)$.

For the thermodynamic analyses of this paper, we can use $K(t)$ rather than $\K(t)$ if the following \textbf{(thermodynamic) closeness} conditions are met:
\begin{enumerate}
\item $K(t)$ obeys SLDB for $\NN^*$, for some set of local temperatures and chemical potentials, and {local} Hamiltonians.
\item $\K(t)$ obeys global LDB, for the same local temperatures and chemical potentials as in (1), for some global Hamiltonian.
\item $\vP(\vx)$ and $\ovP(\vx)$ are extremely close to each other for almost all trajectories $\vx$, e.g., as quantified with KL divergence. 
\item The values of global EP and global EF
that  $\vP(\vx)$ and $\ovP(\vx)$ assign to any trajectory $\vx$ are extremely close to one another, i.e., for any subsystem $i$ and
any transition $x' \rightarrow x : x_{-i} = x'_{-i}$, the global EP and global EF calculated using $K(i;t)$ are close to the global
values calculated using $\K(t)$, as measured on the scale of those global values. Similarly,
for the vector-valued fluctuation theorems to hold, we need the values of the unit EF and unit EP
that $\vP(\vx)$ and $\ovP(\vx)$ assign to any trajectory $\vx$ to be extremely close to one another.

\end{enumerate}
\noindent
The reason that we require that the same local temperatures and chemical potentials occur in closeness conditions (1) and (2)
is that those quantities describe the external reservoirs, and so can be independently measured by the experimentalist;
the goal is that establish that doing the analysis with $K(t)$ rather than $\K(t)$ does not result in experimentally
discernible errors.

As shorthand, I will sometimes simply say that ``$K(t)$ and $\K(t)$ are close''
if they obey the four closeness conditions for some implicit unit structure and initial distribution $p_x(0)$. I will also say that a unit 
graph $\NN^*$ can be \textbf{SLDB-approximated} (typically for an implicit $p_x(0)$) if there is some associated set of choices of $K(t), \K(t)$ such that the four closeness conditions are met.

As an aside, note that if closeness condition (4) holds, and if no work is done on the system, 
then the sum of the actual heat flows into all the subsystems is extremely close to the
sum of the heat flows calculated using the local Hamiltonians. In other words, under these circumstances, we are ensured that energy conservation holds to very
high accuracy, in the sense that $\Delta H_{\vx}$ is very close to the sum over all subsystems $i$ of the
changes in the value of the local Hamiltonian of subsystem $i$, which occur at all moments where $\vx_i$ changes.

Focusing now on just the actual EF, we can 
parallel \cref{eq:19a} to see that for any trajectory $\vx$ and any subsystem $i$, the actual EF into subsystem $i$ from its reservoirs is
\eq{
\oQ^i(\vx) &:=  \sum_{j \in \eta^{-1}(i)}  
\beta_i^{k(j)}\bigg(H_{{\vx}(\tau(j))}(\tau(j)) - H_{{\vx}(\tau(j-1))}(\tau(j))  \nonumber \\
	&\qquad + \mu_{i}^{k(j)}\left[n_i^{k(j)}({\vx}_i(\tau(j-1)) - n_i^{k(j)}({\vx}_i(\tau(j))\right]\bigg) 
\label{eq:d19a}
}
Moreover, because we have a multipartite process, the actual global EF will be a sum of the associated local quantities, 
i.e., 
\eq{
{\overline{Q}}(\vx) = \sum_i {\overline{Q}}^i(\vx)
\label{eq:d19b}
}
(See the definition of local EF
and discussion just before the definition of global EP in the main text.)
In the next subsection I introduce a special type of decomposition of a global Hamiltonian into local Hamiltonians
which reflects the unit structure of the system.
Then in the following subsection I show how this type of global Hamiltonian can be used to SLDB-approximate any given unit structure.

\subsection*{A global Hamiltonian for approximating arbitrary unit structures}

To begin, specify a set of unit-indexed functions $h_{x_\oo}(\oo;t)$. Then
choose the global Hamiltonian to be
\eq{
H_x(t) &= {\sum_{\oo \in {\NN^*}}} h_{x_{\oo}}(\oo;t)
\label{eq:global_Ham}
}
This global Hamiltonian is a sum over units rather than a sum over subsystems.

This type of global Hamiltonian defines global LDB. So to show that a unit structure can be SLDB-approximated with this
type of global Hamiltonian we have to relate it to the local Hamiltonians, which define SLDB.
To do that, recall that SLDB requires that the local Hamiltonian of each subsystem $i$ only
involve $x_{r(i)}$. Accordingly, choose the local Hamiltonian of each subsystem $i$ to be
\eq{
H_{x}(i; t) &= h_{x_{r(i)}}(r(i);t) 
} 

It will be convenient to define a Hamiltonian for each unit $\oo$ which
has the same functional dependence on $h_{\oo'}$ for the
units $\oo' \subseteq \oo$ as the global Hamiltonian, \cref{eq:global_Ham}, has on the units within
$\NN$. This will ensure that even if the full set $\NN$ is a unit,
each unit is ``treated the same'' in the definition of its Hamiltonian.
Accordingly, I define the \textbf{unit Hamiltonian} for each unit $\oo \in \NN^*$ as
\eq{
H_{x_\oo}(\oo; t) &:= \sum_{\oo' \subseteq \oo} h_{x_{\oo'}}(\oo';t)
}
(So in general, $H_x(\oo; t) \ne \sum_{i \in \oo} H_x(i; t)$.)

We can relate this global Hamiltonian and these unit Hamiltonians as follows:
\eq{
H_x(t) = {\widehat{\sum_{\oo \in \NN^*}}} H_{x_\oo}(\oo;t)
\label{eq:global_Ham_in_ex}
}
\begin{proof}
Given the unit structure ${\NN^*}$, define a set $\MMM := \{1, \ldots, |{\NN^*}|\}$. Use that set
to label the units, writing the associated bijection as $c :\MMM \rightarrow {\NN^*}$.
Let $\AAA := \{\AAA_i : i = 1, \ldots |\MMM|\}$ be a unit structure defined over $\MMM$. (So every
$\AAA_i$ is a set of distinct elements from  $\MMM$, there are $|\MMM|$ such sets in $\AAA$, their union
is all of $\MMM$, and they are closed under intersection.) Let $\{g_m(x)\}$ be a set of real-valued
function of $x$, indexed by $m \in \MMM$, and for all $\AAA_i \in \AAA$,
define $g^{\AAA_i }(x) := \sum_{m \in \AAA_i } g_m(x)$. Then by the inclusion-exclusion principle, for all $x$,
\eq{
\sum_{m \in \MMM} g_m(x) = \widehat{\sum_{\alpha \in \AAA}} g^\alpha(x)
\label{eq:d13b}
}

In particular, this is true if for all $i \in \MMM$, $\AAA_i = \cup_{m : c(m) \subseteq c(i)} m$. (Note that $c(m)$ and $c(i)$ are both
subsets of $X$.) With this choice, for all $ i = 1, \ldots |\MMM|$,
\eq{
g^{\AAA_i}(x) = \sum_{m : c(m) \subseteq c(i)} g_{c(m)}(x)}
Choosing $g_{c(m)}(x) =  h_{c(m)}(x)$ for all $m$ and plugging into \cref{eq:d13b} completes the proof.
\end{proof}

Any $\oo \in \NN^*$ defines its own unit structure,
\eq{
\oo^* := \{\oo' \in \NN^* : \oo' \subseteq \oo\}
} 
(Note that $\oo^*$ contains $\oo$ itself, whereas $\NN^*$ need not contain $\NN$.)
\cref{eq:global_Ham_in_ex} applies just as well to any unit $\oo \in \NN^*$, in the sense that
\eq{
H_{x_\oo}(\oo; t) = {\widehat{\sum_{\oo' \in \oo^*}}} H_{x_{\oo'}}(\oo';t)
}


Next, consider a transition  at time $t$, $x' \rightarrow x$, which is mediated by the heat bath of subsystem $i$, so that
$x_{-i} = x'_{-i}$. The actual heat flow into the system from that bath during that transition is
$\beta [H_x(t) - H_{x'}(t)]$, by conservation of energy. In contrast to the subsystem LDB
approximation of the heat that flows in during that transition, $\beta [H_x(i; t) - H_{x'}(i; t)] = \beta [H_{x_{r(i)}}(i; t) - H_{x'_{r(i)}}(i; t)]$, in general
the actual heat that flows in is a function of components of $x'$ and $x$ outside of $r(i)$. (This will be the case whenever there
is a unit that contains $i$ but is not itself a proper subset of $r(i)$.)
More generally, for any trajectory $\vx$, the actual heat flow into subsystem $i$, $\oQ^i(\vx)$,
differs from the subsystem LDB approximation of it,  $Q^i(\vx)$. Similarly, in general the actual heat that flows in
from  the baths of the subsystems in some unit $\oo$, $\oQ^\oo(\vx) = \sum_{i \in \oo} \oQ^i(\vx)$, differs from its subsystem LDB
approximation, $Q^\oo(\vx_\oo)  = \sum_{i \in \oo} Q^i(\vx_{r(i)})$.

Nonetheless, just like its subsystem LDB approximation (see Eq.\,(18) in the main text),
the actual global heat flow is given
by an in-ex sum of the associated unit heat flows:
\eq{
\oQ(\vx) &=  {\widehat{\sum_{\oo \in \NN^*}}} \oQ^\oo(\vx)
}
and similarly, for any unit $\oo \in \NN^*$,
\eq{
\oQ^\oo(\vx) &=  {\widehat{\sum_{\oo' \in \oo^*}}} \oQ^{\oo'}(\vx)
}
So we can define 
\eq{
\oW^\oo(\vx) &:= H_{x_\oo}(\oo;t) - \oQ^\oo(\vx)
\label{eq:actual_unit_work}
}
in order to write the actual global work as 
\eq{
\oW(\vx) &= \Delta H_\vx - \oQ(\vx) \\
	&={\widehat{\sum_\oo}} \oW^\oo(\vx)
}

Similarly, 
we can use \cref{eq:global_Ham_in_ex} to define a (subsystem LDB) \textbf{global work} function by 
\eq{
W(\vx) &:= \Delta H_\vx - Q(\vx) \\
	&= {\widehat{\sum_\oo}} W^\oo(\vx_\oo)
}
where we define the \textbf{unit work} by 
\eq{
W^\oo(\vx_\oo) &:= H_{x_\oo}(\oo;t) - Q^\oo(\vx_\oo)
}
This shows that if we can establish closeness condition (4), so that
the actual EF of a trajectory is close to the associated sum of local EFs,
then actual work done on the system will be close to the SLDB work done 
on the system.

%
%
%
%
%
%

From now on, for simplicity, I assume that there are no energy degeneracies, i.e., there are no two states $x, x' \ne x$
and unit $\oo$ such that $h_x(\oo, t) = h_{x'}(\oo; t)$. (More generally, I assume there are no degeneracies
for any sum of Hamiltonians $h_x(\oo; t)$ that will occur in the analysis.) Also to keep the exposition simple,
I assume that there are no reservoirs connected
to any subsystem except for its heat bath. (So the reservoir indices $k$ will be dropped from now on.) 
Similarly, for most of the rest of this appendix, the time index $t$ is implicit.

\subsection*{Hamiltonian Scaling}

It turns out that any set of rate matrices \{$K(i;t)\}$
can be SLDB-approximated, for appropriate choice of the functions $h(\oo; t)$.
In this subsection I present an example of this. The example is an extreme case,
chosen because it is relatively straightforward; for typical choices of $\{K(i;t)\}$ there is
a much broader set of functions $h(\oo; t)$
that result in the closeness  conditions being met.

To begin, I need to introduce some more notation.
First, create a directed acyclic graph (DAG) $\Gamma = ({\NN^*}, E)$,
where there is an edge $e \in E$ from node $\oo \in \NN^*$ to node $\oo' \in \NN^*$ iff both
$\oo' \subseteq \oo$, and there is no other ``intervening'' unit $\oo''$ such that $\oo' \subseteq \oo'' \subseteq \oo$.
For a unit structure ${\NN^*}$ where $\NN \in {\NN^*}$,
there would be a single root of the DAG $\Gamma$, but in the default case, where  $\NN \not \in {\NN^*}$, $\Gamma$ has
multiple roots. As an example, in Fig.\,(1) in the main text,
$\oo$ and $\alpha$ are the two roots, and $\oo'$ is their (shared)
child. As a notational point, I  will indicate the set of parents of any
node $\oo \in \Gamma$ as $pa(\oo)$, and indicate the set of all of its ancestors as $anc(\oo)$.

%

Note that the state $x_i$ can only occur as an argument of
$h_\oo$ for the unit $\oo = r(i)$ and units $\oo' \in anc(r(i))$. Accordingly,
global LDB requires that for all units $\oo$, subsystems $i \in \oo$, mechanisms $k$, and pairs $x, x' : x_{-i} = x'_{-i}$, the {actual} rate matrix for $i$  obeys
\eq{
\ln \dfrac{\K^{x'}_{x}(i;k)}{\K^{x}_{x'}(i; k)} & = \beta^k_i \bigg(h_{x'_{r(i)}}(r(i)) - h_{x_{r(i)}}(r(i))  \nonumber \\
	&\qquad+ \sum_{\oo' \in anc(r(i))} \left(h_{x'_{\oo'}}(\oo') - h_{x_{\oo'}}(\oo')\right) \nonumber \\
	&\qquad + \mu_{i}^k\left[n_i^k(x_i) - n_i^k(x'_i)\right] \bigg)
\label{eq:actual_heat}
}
(Recall that $t$ is implicit.) In contrast, the rate matrix $K(i; k)$ obeys SLDB, so
\eq{
\ln \dfrac{K^{x'}_{x}(i; k)}{K^{x}_{x'}(i; k)} & =  \beta^k_i \left(h_{x'_{r(i)}}(r(i)) - h_{x_{r(i)}}(r(i)) + \mu_{i}^k\left[n_i^k(x_i) - n_i^k(x'_i)\right]\right) 
\label{eq:fictional_heat}
}
We can combine this with  \cref{eq:19a,eq:d19a} to establish that
the EF part of closeness condition (4) will be met if for all subsystems $i$, for all reservoirs $k$, for all $x_{r(i)}, x'_{r(i)} : x_{-i} = x'_{-i}, x_i \ne x'_i$, 
\eq{
\bigg|\ln \dfrac{\K^{x'}_{x}(i; k)}{\K^{x}_{x'}(i; k)} - \ln \dfrac{K^{x'}_{x}(i; k)}{K^{x}_{x'}(i; k)} \bigg| &< \dfrac{1}{\kappa} \bigg| \ln \dfrac{\K^{x'}_{x}(i; k)}{\K^{x}_{x'}(i; k)} \bigg|
\label{eq:d21}
}
where $\kappa > 0$ is some very large constant. 
By \cref{eq:actual_heat,eq:fictional_heat}, this condition is met if for all such $i, x, x'$,
%
\eq{
\left|h_{x'_{r(i)}}(r(i)) - h_{x_{r(i)}}(r(i))\right| &\ge \kappa \; \bigg|\!\!\! \sum_{\oo' \in anc(r(i))} \left(h_{x'_{\oo'}}(\oo') - h_{x_{\oo'}}(\oo')\right) \bigg|
\label{eq:hamiltonian_ratio_bound}
}
(See App.\,A of~\cite{verley_work_2014}.) 

When \cref{eq:hamiltonian_ratio_bound}
holds for large $\kappa$, I will say that the system obeys \textbf{Hamiltonian scaling}. We have just shown
that the EF part of closeness condition (4) is met for each subsystem if the system obeys Hamiltonian scaling.
In the rest of this appendix I show that the rest of the closeness conditions also hold under Hamiltonian scaling.

From now on, for simplicity I assume there are no chemical reservoirs, and that $\beta_i = 1$ for all $i$.
I also make $t$ explicit again, temporarily, and adopt the common notation that ``$diag({\vec{a}})$''
means the matrix that has zeros on all entries except the diagonal, which has the entries of $\vec{a}$.
Recall that we can write a rate matrix $K(i; t)$ that obeys SLDB  as
\eq{
K(i;t) &= R(t) \Pi (i;t)- {\mbox{diag}}\left(R(t) \vec{\pi}(i; t)\right)
\label{eq:d24}
}
where $R(t)$ is an arbitrary symmetric matrix;
\eq{
\pi_x(i;t) &:= \dfrac{e^{- H_x(i;t)}}{Z(i;t)}
\label{eq:d23b}
}
is the equilibrium distribution of the local Hamiltonian $H_x(i;t)$; $\Pi(i; t)$ is the diagonal
matrix with $\pi_x(i; t)$ on the diagonal; and $\vec{\pi}(i; t)$ is that distribution expressed as a column vector.
Since the process is multipartite, we can take $R^{x'}_x(t) = 0$ for all $x', x$ that differ in 
two or more components.

Similarly, we can choose the rate matrix $\K(i; t)$ that obeys global LDB  so that
\eq{
\K(i; t) &= R(t) \oPi (t)- {\mbox{diag}}\left(R(t) \vec{\opi}(t)\right)
\label{eq:d23}
}
where 
\eq{
\opi_x(t) &:= \dfrac{e^{- H_x(t)}}{\oZ(t)}
\label{eq:d24b}
}
is the equilibrium distribution of $H_x(t)$ for fluctuations arising from the heat bath of subsystem $i$; $\oPi(t)$ is the diagonal
matrix with $\opi_x(t)$ on the diagonal; and $\vec{\opi}(t)$ is that distribution expressed as a column vector.
(Note that since $\beta_i$ is the same for all subsystems, $\K(i; t)$ is independent of $i$.)

Recall that the actual probability density function evaluated at any trajectory $\vx$ is~\cite{esposito2010threedetailed}
\eq{
\ovP(\vx) &= p_{\vx(0)}(0) \prod_{j=1}^{C(\vx)}  \left[\K^{\vx(\tau(j-1))}_{\vx(\tau(j))}(\tau(j))  e^{{}^{-\int_{\tau(j-1)}^{\tau(j)} dt \, \K^{\vx(\tau(j-1))}_{\vx(\tau(j-1))}(t)}} \right]
}
and similarly for $\vP(\vx)$. Therefore the log-ratio of the actual and SLDB probability density functions evaluated at $\vx$ is 
\eq{
\ln \left[\dfrac{\ovP(\vx) }{\vP(\vx) } \right] &= \sum_{j=1}^{C(\vx)} \ln\left[  \dfrac{\K^{\vx(\tau(j-1))}_{\vx(\tau(j))}(\tau(j))  e^{{}^{-\int_{\tau(j-1)}^{\tau(j)} dt \, \K^{\vx(\tau(j-1))}_{\vx(\tau(j-1))}(t)}} }
				{K^{\vx(\tau(j-1))}_{\vx(\tau(j))}(\tau(j))  e^{{}^{-\int_{\tau(j-1)}^{\tau(j)} dt \, K^{\vx(\tau(j-1))}_{\vx(\tau(j-1))}(t)}} }  \right]   \\
	&= \sum_{j=1}^{C(\vx)} \ln\left[  \dfrac{\K^{\vx(\tau(j-1))}_{\vx(\tau(j))}(\tau(j))} 	{K^{\vx(\tau(j-1))}_{\vx(\tau(j))}(\tau(j)) }\right] \nonumber \\
	& \qquad +\int_{\tau(j-1)}^{\tau(j)} dt  \left[ K^{\vx(\tau(j-1))}_{\vx(\tau(j-1))}(t)  - \K^{\vx(\tau(j-1))}_{\vx(\tau(j-1))}(t) \right]
\label{eq:d29}
}


To evaluate \cref{eq:d29}, pick any $j \in \{1, \ldots, C(\vx)\}$ and as shorthand write 
$i = \eta(j)$, $x = \vx(\tau(j))$, and $x^{-} = \vx(\tau(j-1))$.
Since $x^-_i \ne x_i$, ${\mbox{diag}}\left(R(t) \vec{\pi}(i; t)\right)^{x^-}_x = {\mbox{diag}}\left(R(t) \vec{\opi}(t)\right)^{x^-}_x = 0$. 
Since the state change at $\tau(j)$ is due to an energy transfer between the system and the heat reservoir of subsystem $i$,
the term in the sum on the RHS of \cref{eq:d29} corresponding to this choice of $j$ is
\eq{
 \ln \left[ \dfrac{ \K^{x^-}_x(i; t)} {K^{x^-}_x(i;t) }\right] &=  \ln \left[ \dfrac{ \left(R(t) \oPi (t)\right) |^{x^-}_x} {\left(R(t) \Pi (i; t)\right)  |^{x^-}_x}\right] \\
	&= \ln \left[ \dfrac{ R(t)^{x^-}_x \oPi (t)^x_x} {R(t)^{x^-}_x \Pi (i; t)^x_x} \right] \\
	&= \ln \left[ \dfrac{ \opi (t)_x} {\pi (i; t)_x}\right]
\label{eq:d34}
}
(Note that the LHS is independent of $x^-$, in contrast to the LHS.)

Next, note that when Hamiltonian scaling holds, we can rearrange the numerator and denominator terms in \cref{eq:d21} and then plug in \cref{eq:d34}
to show that for all subsystems $i$, for all $x_{r(i)}, x'_{r(i)} : x_{-i} = x'_{-i}, x_i \ne x'_i$, 
\eq{
\bigg| \ln \left[ \dfrac{ \opi (t)_x} {\pi (i; t)_x}\right] - \ln \left[ \dfrac{ \opi (t)_{x'}} {\pi (i; t)_{x'}}\right] \bigg|
 &< \frac{1}{\kappa} \bigg| \ln \left[ \dfrac{ \opi (t)_{x'}} {\opi (t)_x}\right] \bigg|
}
Therefore for all $x_{-i}$,
\eq{
&\!\!\!\!\!\!\!\! \max_{x_i, x_{-i}} \, \ln \left[ \dfrac{ \opi (t)_{x_i, x_{-i}}} {\pi (i; t)_{x_i, x_{-i}}}\right] \;-\; \min_{x'_i} \, \ln \left[ \dfrac{ \opi (t)_{x'_i, x_{-i}}} {\pi (i; t)_{x'_i, x_{-i}}}\right]   \nonumber \\
 &\qquad\qquad\qquad<  \frac{1}{\kappa}  \max_{x'_i,x_i} \, \ln \left[ \dfrac{ \opi (t)_{x'_i, x_{-i}}} {\opi (t)_{x_i, x_{-i}}}\right] 
\label{eq:d34a}
}

Next, expand the logarithm on the RHS of \cref{eq:d34a} as
\eq{
& h_{x_{r(i)}}(r(i)) - h_{x'_{r(i)}}(r(i)) + \!\!\!\!\!  \sum_{\oo' \in anc(r(i))} \left(h_{x_{\oo'}}(\oo') - h_{x'_{\oo'}}(\oo')\right) \\
& \qquad \simeq  h_{x_{r(i)}}(r(i)) - h_{x'_{r(i)}}(r(i)) 
}
where the second line follows from the assumption of Hamiltonian scaling. Therefore \cref{eq:d34a} means that for any fixed scale of $h(r(i))$, i.e., any fixed value of
\eq{
 \max_{x'_i,x_i}    \left[h_{x'_{i}, x_{r(i) \setminus i}}(r(i)) - h_{x_{i}, x_{r(i) \setminus i}}(r(i)) \right] \;,
\label{eq:d35a}
}
by taking $\kappa$ large enough we can make $ \ln \left[ \dfrac{ \opi (t)_x} {\pi (i; t)_x}\right]$ be arbitrarily close to $0$
for all $x$.  Combining this with \cref{eq:d34} means that the sum on
the RHS of \cref{eq:d29} is arbitrarily close to $0$, for large enough $\kappa$.

Next, for all $j, t \in [\tau(j), \tau(j+1))$, expand the escape rate
under SLDB as the sum of the escape rates due to fluctuations generated by any of the subsystem's reservoirs:
\eq{
-K^{\vx(\tau(j))}_{\vx(\tau(j))}(t) &= -K^{x}_{x}(t)  \\
		&= - \sum_{i} K^{x}_{x}(i; t) \\
	&= - \sum_{i} \left( {\mbox{diag}}\left(R(t) \vec{\pi}(i; t)\right) \right)^{x}_{x} \\
	&= - \sum_{i}  \sum_{x'} R(t)^{x}_{x'} \pi_{x'}(i; t)
}
Similarly, to calculate the escape rate for $\K^{\vx(\tau(j))}_{\vx(\tau(j))}(t)$ we need to sum over all
the reservoirs. In this case though, the equilibrium distribution is independent of $i$. Accordingly,
\eq{
-\K^{\vx(\tau(j))}_{\vx(\tau(j))}(t) &= - \sum_{i}  \sum_{x'} R(t)^{x}_{x'} \opi_{x'}(t)
}

Combining, and using the fact that $R^{x'}_x(t) = 0$ if $x'$ and $x$ differ in two more components,
\eq{
& \int_{\tau(j-1)}^{\tau(j)} dt  \left[ K^{\vx(\tau(j))}_{\vx(\tau(j))}(t)  - \K^{\vx(\tau(j))}_{\vx(\tau(j))}(t) \right]   \nonumber \\
	&\qquad =  \sum_{i,x'_{i}}
			\int_{\tau(j-1)}^{\tau(j)} dt  \, R(t)^{x^-_{i}, x^{-}_{-i}}_{x'_i,  x^{-}_{-i}} \left[ \opi(t)_{x'_i,  x^{-}_{-i}} - \pi(i';t)_{x'_i,  x^{-}_{-i}}  \right]
\label{eq:d35}
}
As shown by the discussion surrounding \cref{eq:d35a}, under Hamiltonian scaling, 
$\opi(t)_{x'_i,  x^{-}_{-i}} - \pi(i';t)_{x'_i,  x^{-}_{-i}}$ can be made arbitrarily close to $0$ by choosing large enough $\kappa$.
Therefore the integral on the RHS of \cref{eq:d29} is also arbitrarily close to $0$ under Hamiltonian scaling.

Finally, suppose that the probability under $\K(t)$ of a trajectory $\vx$ such that $C(\vx) = 0$  is infinitesimally
small, i.e., that $\ovP(\vx)$ of such a trajectory is close to zero. Then combining the results just above,
we see that under Hamiltonian scaling, for large enough $\kappa$,
\eq{
\int d\vx \ovP(\vx) \ln \left[\dfrac{\ovP(\vx)}{\vP(\vx)}\right]
}
is arbitrarily close to $0$. This means that closeness condition (3) is met if Hamiltonian scaling holds
(and the probability under $\K(t)$ of a trajectory $\vx$ such that $C(\vx) = 0$  is infinitesimally
small).

The only remaining condition to establish is the EP part of closeness condition (4). 
Since $K$ and $\K$ assign almost equal probabilities to almost all trajectories $\vx$ under Hamiltonian scaling ,
they also assign almost equal values to $\Delta s_i(\vx)$. Moreover, I showed above that under
Hamiltonian scaling, $K$ and $\K$ assign almost equal values to the local EFs of $\vx$, i.e., for all subsystems $i$, $Q^i(\vx) = \oQ^i(\vx)$ to high accuracy. 
Given the definition of global EP,
this establishes that
under Hamiltonian scaling, $K$ and $\K$ assign almost equal values to the global and unit EPs of $\vx$. 
This establishes that closeness condition (4) holds in full under Hamiltonian scaling. 

Combining, we see that under Hamiltonian scaling for large enough $\kappa$, 
all four closeness conditions can be met. Note that no assumption concerning the form of the unit structure
is needed to get this result. So as claimed, any unit structure $\NN^*$ can be SLDB-approximated.



\bibliographystyle{ws-rv-van}
\bibliography{../../../../../BIB/thermo_refs.main.3.BIB.DIR}

\end{document}